\documentclass[12pt]{article}

\usepackage{amssymb}
\usepackage{amsmath}
\usepackage{bm}
\usepackage{bbm}
\usepackage{array,multirow}
\usepackage{hyperref}
\usepackage{enumerate}
\usepackage{enumitem}
\usepackage{longtable}
\usepackage[super,numbers]{natbib}
\usepackage{graphicx}
\usepackage{geometry}
\usepackage{rotating}

\usepackage{setspace}
\usepackage{pdflscape}

\newcommand{\beq}{\begin{equation}}
\newcommand{\eeq}{\end{equation}}
\newcommand{\bei}{\begin{itemize}}
\newcommand{\eei}{\end{itemize}}
\newcommand{\ben}{\begin{enumerate}}
\newcommand{\een}{\end{enumerate}}

\newcommand{\E}{{\rm E}}
\newcommand{\V}{{\rm V}}

\newcommand{\logit}{{\rm logit}}
\newcommand{\expit}{{\rm expit}}
\newcommand{\argmin}{{\rm arg\,min}}

\begin{document}

\title{Confounder selection strategies targeting stable treatment effect estimators}

\author{
Wen Wei Loh\footnote{Department of Data Analysis, Ghent University, Gent, Belgium}, and 
Stijn Vansteelandt\footnote{Department of Applied Mathematics, Computer Science and Statistics, Ghent University, Ghent, Belgium} \footnote{Department of Medical Statistics, London School of Hygiene and Tropical Medicine, United Kingdom}
}

\maketitle

\begin{abstract}
Inferring the causal effect of a treatment on an outcome in an observational study requires adjusting for observed baseline confounders to avoid bias. However, adjusting for all observed baseline covariates, when only a subset are confounders of the effect of interest, is known to yield potentially inefficient and unstable estimators of the treatment effect. Furthermore, it raises the risk of finite-sample bias and bias due to model misspecification. For these stated reasons, confounder (or covariate) selection is commonly used to determine a subset of the available covariates that is sufficient for confounding adjustment. 
In this article, we propose a confounder selection strategy that focuses on stable estimation of the treatment effect. In particular, when the propensity score model already includes covariates that are sufficient to adjust for confounding, then the addition of covariates that are associated with either treatment or outcome alone, but not both, should not systematically change the effect estimator. 
The proposal, therefore, entails first prioritizing covariates for inclusion in the propensity score model, then using a change-in-estimate approach to select the smallest adjustment set that yields a stable effect estimate. 
The ability of the proposal to correctly select confounders, and to ensure valid inference of the treatment effect following data-driven covariate selection, is assessed empirically and compared with existing methods using simulation studies. We demonstrate the procedure using three different publicly available datasets commonly used for causal inference.
\end{abstract}

\noindent%

\section{Introduction}

Inferring the causal effect of a treatment (or an exposure) on an outcome in an observational study requires adjusting (or controlling) for observed baseline confounders to avoid bias. 
In practice, multiple (continuous) covariates are usually available, and it is not a priori known which of these truly confound the causal effect of interest. Adjusting for all the observed covariates is often not possible in view of the 
curse of dimensionality, and even if possible, may imply an unwarranted loss of statistical power, and moreover induce finite-sample bias and bias due to model misspecification \citep{greenland2008invited,stephens2013flexible}. Confounder (or covariate) selection procedures are commonly employed to prevent this. 
Recently, Witte and Didelez\cite{witte2018covariate} proposed a classification scheme for such procedures based on the type of target adjustment set and selection mechanism. In Section~\ref{sect:review} we will review some of these procedures and highlight a number of shortcomings as motivation for the proposed strategy, which we introduce in Section~\ref{sect:intro_proposed}.

\subsection{Existing covariate selection approaches \label{sect:review}}

{\it Propensity score} (PS) methods \citep{rosenbaum1983central,rosenbaum1987model} are widely adopted in observational studies to adjust for confounding, by summarizing the multiple (continuous) covariates in terms of the conditional probability of treatment given observed baseline covariates.
When selecting covariates for inclusion in the PS model, the conventional `design of observational studies' typically suggests that only the treatment and covariates be used without access to the outcome\citep{stuart2010matching,rubin2007design,imbens2015causal}. 
However, a limitation of such approaches is that the predictive ability and significance of covariates in the PS model are not directly informative about the extent to which confounding bias is reduced\citep{vanderweele2019principles}, and ultimately, the quality of the treatment effect estimator. Methods that test only the covariate-treatment associations can moreover worsen variance inflation by selecting non-confounding but strong predictors of treatment \citep{brookhart2006variable,vansteelandt2012model}. 
This has stimulated recent development of approaches, such as collaborative targeted minimum loss-based estimation\citep{van2010collaborative}, outcome-adaptive LASSO\citep{shortreed2017outcome}, and Bayesian Adjustment for Confounding (BAC) algorithm\citep{wang2015accounting}, among others, which seek to penalize covariates that predict treatment only in the PS model, albeit at the risk of invalidating inference for the treatment effect. 
A different approach is to directly evaluate the impact of covariate selection on the treatment effect estimator by adopting a `change-in-estimate' (CIE) perspective \citep{maldonado1993simulation,mickey1989impact}. For example, a sequence of nested covariate subsets (ranging from the empty set with no covariates to the full set containing all covariates) is constructed, and the effect estimates that adjust for the covariates in each subset calculated. 
The (smallest) covariate subset whose effect estimate remains unchanged (within some pre-specified threshold, e.g., 10\% of the absolute magnitude) after controlling for additional covariates is then selected for inference on the treatment effect.
Greenland et al\cite{greenland2016outcome}, and Vansteelandt et al\cite{vansteelandt2012model}, improve on such a procedure: they optimize the quality of the effect estimator by considering changes in the approximate mean-squared error (MSE) of the conditional effect estimator (based on an outcome regression model that includes treatment and the covariates in each subset), and doubly-robust marginal effect estimator (based on an outcome model and a treatment model that includes the covariates in each subset), respectively.
While these approaches offer indirect insight into the sensitivity of the treatment effect estimator to the selected covariate subset, they do not fully exploit its insensitivity with the aim to achieve approximately valid inference following such a data-driven confounder selection procedure.

\subsection{Selecting confounders based on the stability of the treatment effect estimator \label{sect:intro_proposed}}

In view of this shortcoming, in this article, we propose a confounder selection strategy that entails first {\it prioritizing} covariates for confounding adjustment, then extending the CIE approach to select the smallest subset of covariates that yields a {\it stable} effect estimate. 
In the first part, the covariates are ordered by decreasing priority for confounding adjustment using a forward selection approach. The resulting sequence therefore indexes a series of nested covariate subsets: the smallest non-empty subset contains the most important covariate (according to a specified criterion to be defined), the next smallest (non-empty) subset contains the two most important covariates, and so on, with the largest subset containing all the covariates. We elaborate on the detailed steps in Section~\ref{sect:part1_prioritize}.
Explicitly ordering the covariates permits researchers to use substantive knowledge to refine further which covariates should have higher or lower priorities for confounding adjustment (in the PS model). 
Existing matching methods such as Zubizarreta\cite{zubizarreta2012using}, among others, already allow users to incorporate substantive expertise in guiding the covariate adjustment process.
Keele and Small\cite{keele2018comparing} recently compared matching methods that allow for covariate prioritization with machine learning methods that require little user input toward obviating biases due to observed confounding.

In the second part, the {\it stability} of the effect estimator across the series of nested covariate subsets indexed by the prioritized covariates is assessed.
Suppose that the available set of covariates contains a subset that is sufficient to adjust for confounding. Intuitively, the standardized treatment effect estimator that already adjusts for the confounders in this sufficient subset should remain relatively steady, even after further adjusting for other non-confounders, i.e., those associated with either treatment or outcome, but not both.
Most covariate selection strategies account for such relative insensitivity only to the extent of recommending in practice that a given procedure merely be repeated for different subsets as a form of post-hoc sensitivity analysis.
In contrast, in this article, we propose a selection strategy that works along a different principle than routine methods, by exploiting this knowledge to explicitly assess the trajectory of the treatment effect estimator across different (nested) covariate subsets.
The smallest subset that yields an unaltered estimator (as more covariates are added), relative to a `benchmark' estimate (e.g., one that adjusts for all covariates), is selected. We elaborate on the detailed steps in Section~\ref{sect:part2_stability}.
{\it Randomization(-based)} inference \citep{Rosenbaum:2002} about the causal effect can then be carried out, by comparing treated and untreated individuals within strata constructed by {\it matching on the PS}\citep{stuart2010matching} containing only the selected covariates. A review of randomization inference is provided in Section~\ref{sect:reviewRI}. 
While valid inference cannot be guaranteed following a data-driven selection of covariates, we argue that nearly valid inference using the strategy proposed in this article may be possible with the combined use of (i) double selection for prioritizing the covariates, (ii) stability-based assessment to select covariates for confounding adjustment, and (iii) randomization inference using PS matching to control the type I error when testing the null hypothesis of no (individual) treatment effect.

The rest of the article is as follows. In Section~\ref{sect:select}, the proposed two-part confounder selection strategy is presented. In Section~\ref{sect:sims}, the ability of the proposed procedure to correctly select confounders of the treatment-outcome relation, and to control the type I error rate when testing the null of no individual treatment effect following covariate selection, is evaluated empirically in simulation studies. The performance of the proposal is compared with existing methods for covariate selection, and for inference on the treatment effect following covariate selection.
In Section~\ref{sect:examples}, the proposed procedure is demonstrated using three different publicly available datasets commonly used for causal inference. We conclude with a brief discussion in Section~\ref{sect:discuss}.

\section{A new proposal for confounder selection based on stability of the treatment effect estimator}\label{sect:select}

\subsection{Confounder selection procedure}
In this section we describe the proposed confounder selection procedure which proceeds in two parts. 
\begin{enumerate}
\item Order the covariates, e.g., based on the strength of their (conditional) associations with treatment and with outcome within each orbit, in decreasing priority for confounding adjustment. 
\item For the series of nested covariate subsets induced by the ordered covariates, calculate the treatment effect estimator for each orbit. Select the covariate subset corresponding to the smallest orbit that yields the most stable effect estimator.
\end{enumerate}

In the first part, the covariates are ordered to reflect their (decreasing) priority for confounding adjustment. 
To facilitate partitioning the space of all possible covariate subsets and visualizing the resulting nested covariate subsets, we exploit the {\it orbits} framework of Crainiceanu et al\cite{crainiceanu2008adjustment}. In particular, we define the $j$-th orbit to comprise all subsets with $j+1$ covariates, including an intercept. Orbits may be more generally used to select covariates for confounding adjustment, such as via regression adjustment in the outcome model alone, or via inclusion in the outcome and treatment models when calculating doubly-robust (marginal) treatment effect estimators; see e.g.,  Vansteelandt et al\cite{vansteelandt2012model}. However, in this article, we will focus on covariate selection for the PS model to be used for matching in constructing strata for randomization inference.
Each candidate covariate is evaluated, in turn, following {\it double selection} \citep{belloni2014inference} principles, where the (partial) associations between the outcome and the covariates, and between the treatment and the covariates, are considered.
Using double selection can increase the likelihood of selecting true confounders, and account for the variability induced by carrying out covariate selection on the same data.

\subsubsection{Prioritizing covariates for confounding adjustment \label{sect:part1_prioritize}}
Let $J$ denote the total number of observed covariates so that there are $J+1$ different orbits. 
In this article, we assume that there is a sufficient number of observations so that the regression model for the treatment (outcome) containing all $J$ observed covariates (and treatment) can be fitted to the observed data.
We further assume that {\it conditional exchangeability} holds given all the observed covariates; i.e., that each individual's probability of being assigned to treatment does not depend on the potential outcomes or any other (unobserved) variables.
%We will henceforth refer to a particular covariate subset, and the PS model that contains said covariates in the subset, interchangeably.
The first part then proceeds by repeating the following steps for each orbit indexed by $j=1,\ldots,J$.
\begin{enumerate}
\item Let $L^{j-1}$ denote the subset of covariates selected from the $(j-1)$-th orbit. When $j=1$, let $L^{0}$ denote the set containing the constant (intercept) $1$ only. Denote each of the remaining covariates that is not in the subset $L^{j-1}$ by $L^{j-1, k}, k=1,\ldots,J-(j-1)$.
These $J-(j-1)$ covariates are therefore candidates for selection in the $j$-th orbit. 
\item 
For $k=1,\ldots,J-(j-1)$, evaluate the (conditional) association between each candidate covariate $L^{j-1, k}$ and the treatment $A$ by fitting the regression model for the treatment:
\beq\label{eq:ds_PSmodel}
g\{\E(A|L^{j-1}, L^{j-1, k})\}= \alpha_{j-1} L^{j-1} + \alpha_{j-1,k} L^{j-1, k},
\eeq
where $\alpha_{j-1}$ is a coefficient vector of length $j$ (including the intercept), and $g$ is a link function such as the $\logit$ link, $\logit(x)=\log\{x/(1-x)\}$. Similarly, fit the regression model for the outcome $Y$: 
\beq\label{eq:ds_Ymodel}
h\{\E(Y|A,L^{j-1}, L^{j-1, k})\} = \psi_{j,k} A + \beta_{j-1} L^{j-1} + \beta_{j-1,k} L^{j-1, k},
\eeq
where $\beta_{j-1}$ is a coefficient vector of length $j$ (including the intercept), and $h$ is a link function such as the identity link, $h(x)=x$.
Non-continuous treatments, or outcomes, can thus be accommodated using non-linear models with non-identity links $g(\cdot)$, or $h(\cdot)$, respectively. 
The subscripts in the treatment coefficient $\psi_{j,k}$ denote the dependence on the covariates being adjusted for in the $j$-th orbit, $(L^{j-1}, L^{j-1, k})$.
The regression models for the treatment and outcome can therefore be used to compare the relative strengths of the covariate-treatment and covariate-outcome (partial) associations, encoded by $\alpha_{j-1,k}$ and $\beta_{j-1,k}$ respectively, across all candidate covariates $L^{j-1, k}, k=1,\ldots,J-(j-1)$ that remained unselected in the subset $L^{j-1}$ from the $(j-1)$-th orbit. 

\item Test the hypotheses $H_0(A,L^{j-1, k}):\alpha_{j-1,k}=0$ and $H_0(Y,L^{j-1, k}):\beta_{j-1,k}=0$ separately using e.g., the Wald test. Denote the resulting p-values by $pv(A,L^{j-1, k})$ and $pv(Y,L^{j-1, k})$ respectively. Let $k^\ast$ denote the index of the candidate covariate that minimizes the minimum of the p-values $pv(A,L^{j-1, k})$ and $pv(Y,L^{j-1, k})$; i.e.,
\beq\label{eq:forwardselect_minimin}
k^\ast = \underset{k}{\argmin} \left\{ \min \left( pv(A,L^{j-1, k}),pv(Y,L^{j-1, k})\right) \right\}.
\eeq
Define the selected subset for the $j$-th orbit to be $L^j=L^{j-1} \cup L^{j-1, k^\ast}$.
Denote the estimated (conditional) average treatment effect, given the selected covariates $L^j$ in the fitted outcome regression model, by $\widehat \psi_{j} = \widehat \psi_{j,k^\ast}$.
\end{enumerate}
Repeating the steps above for $j=1,\ldots,J$, therefore returns a sequence of covariates, $L^{0, k^\ast}, \ldots, L^{J-1, k^\ast}$, that are ordered by their decreasing priority for confounding adjustment. In particular, the ordered covariates induce a series of nested covariate subsets, $L^1 \supset \ldots \supset L^{J}$.

Explicitly prioritizing the covariates for confounding adjustment enables further refinement by leveraging established substantive knowledge. For example, covariates that are known common causes of the treatment-outcome relation may be promoted to higher priorities for adjustment by the researcher, whereas known instrumental variables (or ``instruments'') that only affect the treatment may be relegated to lower priorities. 
In general, determining a variable to be an instrument requires (scientific) knowledge that the purported instrument causes the treatment, but is entirely unassociated with the outcome except possibly via treatment. Such considerable knowledge may be used to justify stringent assumptions for instrumental variable analyses in certain contexts, but is rarely available in most substantive analyses.
Covariates that are a priori known to be correlated may be grouped together (e.g., by ordering them consecutively in practice); however, adjusting for one confounder that is correlated with another can (partially) eliminate the biases that may arise when the latter is unadjusted for \citep{stuart2010matching}.
Taking the minimum of the p-values for the covariate-treatment and covariate-outcome associations in \eqref{eq:forwardselect_minimin} is inspired by double selection principles that take advantage of information on both these associations.
Assessing the association between each candidate confounder with only the outcome (or treatment), instead of simultaneously as in \eqref{eq:forwardselect_minimin}, may potentially increase the chances of detecting variables that more strongly predict outcome (or treatment). However, the associations between the covariates with either treatment or outcome is then entirely ignored, which can potentially discard confounders that are common causes of both treatment and outcome, but more strongly associated with one than the other.
Using double selection principles therefore reduces the risk of biased inference from failing to adjust for confounders, while possibly producing more variable estimators, but which arguably better reflects the inherent uncertainty in the data.
Notwithstanding the different orderings that may potentially result from different measures for prioritizing the covariates, no covariates are precluded from confounding adjustment following this first part of the procedure, because no pre-determined (significance-based) threshold is imposed to eliminate covariates based solely on the absence of association.

In principle, the series of nested covariate subsets may be constructed using a backward or stepwise selection approach; e.g., starting with the full model, the candidate covariate most weakly associated with either treatment or outcome (e.g., the coefficient with the largest p-value across the treatment and outcome regression models) can be iteratively discarded from the selected subset in each orbit. 
However, a forward selection approach has the benefit of avoiding convergence issues that may potentially arise when fitting regression models with many covariates.
When the number of covariates further exceeds the number of observations, (unpenalized) regression models for the treatment and the outcome can theoretically be fitted in orbits with fewer covariates than observations. 
Extensions to settings with high-dimensional observed covariates, which require either limiting the largest orbit to a selected covariate subset with fewer covariates than observations, or introducing penalized regression models in larger orbits, are beyond the scope of this article and deferred to future work.

\subsubsection{Stability of the effect estimator \label{sect:part2_stability}}

In the second part of the proposed strategy, the smallest subset of covariates is selected for confounding adjustment by assessing the stability of the treatment effect estimator across the nested covariate subsets. 
In particular, common covariate selection strategies predominantly identify a sufficient subset of covariates for confounding adjustment by optimizing a stopping criterion through separate evaluations of each subset. 
In most realistic settings, it is impossible to determine whether the single selected subset that met the stopping criterion is sufficient to eliminate all biases due to observed (baseline) confounding. 
When there exists a subset of covariates that is sufficient to control for confounding, further adjusting for other covariates that are associated with either treatment or outcome, but not both, should not systematically change the treatment effect estimator.
We therefore exploit this knowledge by explicitly evaluating the trajectory of the treatment effect estimator across the (nested) covariate subsets induced by the prioritized covariates in the first part. 

We briefly introduce the notation before presenting the effect estimator for evaluating stability.
In a sample of size $n$, for individual $i=1,\ldots,n$, denote the binary treatment by $A_i$ and the outcome of interest by $Y_i$.
Let $Y^a_i$ denote the potential outcome for $Y_i$ if, possibly counter to fact, individual $i$ had been assigned to treatment level $A_i=a$. In this article, our interest is in the marginal treatment effect, defined as $\psi=\E(Y^1)-\E(Y^0)$. 
We propose comparing the treatment effect estimator from each orbit, $\widehat \psi_{j}, j=1,\ldots,J-1$, with the `benchmark' estimator $\widehat \psi_{J}$ that adjusts for all available covariates (i.e., from the largest orbit), though other benchmarks can be readily considered in general. 
The choice of $\widehat \psi_{J}$ as a reference is merely motivated by its (asymptotic) unbiasedness assuming that there are no unobserved confounders, so that the difference $\widehat\psi_{j}-\widehat\psi_{J}$ can be viewed as an approximate bias of $\widehat\psi_{j}$ relative to the true value of $\psi$.
In this article, we will use the standardized difference between the treatment effect estimators to evaluate their stability across the different orbits, which takes the form:
\beq\label{eq:score_diff_std_fullmodel}
\check\psi_{j} = \dfrac{\widehat\psi_{j}-\widehat\psi_{J}}
{\sqrt{\V\left(\widehat\psi_{j}-\widehat\psi_{J}\right)}},
\quad j=1,\ldots,J-1;
\eeq
where $\V(X)$ denotes the asymptotic variance of $X$.
Our goal now is not to merely select the orbit that minimizes the absolute value of \eqref{eq:score_diff_std_fullmodel}, but to select the smallest orbit that yields the most `stable' value of \eqref{eq:score_diff_std_fullmodel}.
For example, the values of \eqref{eq:score_diff_std_fullmodel} may be plotted for each orbit, as we will illustrate using the applied examples later, and the (smallest) orbit that yields a value that does not `fluctuate' as the orbits grow can be selected through visual inspection.

Alternatively, a diagnostic that numerically assesses the relative stability or (in)sensitivity of \eqref{eq:score_diff_std_fullmodel}, while taking into account its variability, can directly optimize the quality of the treatment effect estimator.
In this article, we will use an inverse variance weighted average of the differences $\widehat\psi_{j}-\widehat\psi_{J}$ within a (moving) window of consecutive orbits. For simplicity, we will use (symmetric) windows of width five centered around each orbit $j=3,\ldots,J-2$. The diagnostic for the $j$-th orbit is therefore defined as:
\beq\label{eq:Q_diff_std_fullmodel}
Q_j = \sum_{k=j-2}^{j+2} w_k \{(\widehat\psi_{k}-\widehat\psi_{J}) - \overline{\widehat\psi_{j}}\}^2,
\eeq
where the weights $w_k$ and weighted average $\overline{\widehat\psi_{j}}$ are respectively defined as:
\[
w_k = \left\{\V\left(\widehat\psi_{k}-\widehat\psi_{J}\right)\right\}^{-1},
\quad
\overline{\widehat\psi_{j}} = \left(\sum_{k=j-2}^{j+2} w_k\right)^{-1}\sum_{k=j-2}^{j+2} w_k (\widehat\psi_{k}-\widehat\psi_{J}).
\]
It is hence assumed that $w_J=0$. 
Notwithstanding the $\widehat\psi_{J}$ estimators canceling each other out within the quadratic term in \eqref{eq:Q_diff_std_fullmodel}, we use weights that are inversely proportional to the variance of the approximate biases encoded by the differences $\widehat\psi_{j}-\widehat\psi_{J}$ toward attaining unbiased inference, rather than the variance of $\widehat\psi_{j}$ alone which focuses on the efficiency of the effect estimator.
Correlations between (non-)consecutive orbits are accounted for through the dependence on the common benchmark estimator $\widehat \psi_{J}$. 
The differences between the treatment effect estimator $\widehat \psi_{j}$ from orbit $j$ and the benchmark estimator $\widehat \psi_{J}$, and the associated variances, can be consistently estimated under settings with (non-)linear parametric regression models for the treatment and the outcome as described later. The smallest orbit with the most stable value of \eqref{eq:score_diff_std_fullmodel} therefore minimizes the Q statistic; i.e.,
\beq\label{eq:diagnostic_min_QJ}
\min_{j=2,\ldots,J-1} \; Q_j.
\eeq
The weighted average as defined in \eqref{eq:Q_diff_std_fullmodel} adopts the same form as `Cochran's Q statistic'\citep{hoaglin2016misunderstandings} from the meta-analysis literature for assessing heterogeneity of effect-size estimates from separate studies. Because the Q statistic is used merely to summarize the (in)stability of the effect estimators across different orbits, and not compared against its asymptotic distribution for inference, no assumption of independence between (non-)consecutive orbits is needed.

In general, the conditional effect estimator encoded by the treatment coefficient in the (possibly non-linear) outcome model \eqref{eq:ds_Ymodel} may suffer from non-collapsibility\citep{greenland1999confounding}, and therefore lack any stable behavior as additional covariates are adjusted for, even when they are non-confounders. For instance, even when all the (true) confounders have been adjusted for, further adjustment for covariates that are associated with outcome (only) can change the estimated odds ratio, and potentially result in an artificial lack of perceived stability.
To facilitate assessing stability across different orbits when evaluating non-collapsible effect measures, we recommend using a marginal effect estimator based on doubly robust standardization\citep{vansteelandt2011invited}.
This approach delivers an unbiased estimator if either the outcome or the treatment model is correctly specified, without amplifying biases that may arise due to the misspecified model. Furthermore, this approach employs standardization using traditional outcome regression, thereby inheriting the benefit of increased precision but not the risks of biases due to extrapolation.
Lastly, the (asymptotic) variance estimator in \eqref{eq:score_diff_std_fullmodel} can be derived in closed form for computational efficiency.
The estimator is calculated as follows.
For a binary treatment, denote the (non-linear) treatment model in \eqref{eq:ds_PSmodel}, conditional on the selected covariates in the $j$-th orbit $L^{j}$, by $\E(A_i|L^{j}_i)=\Pr(A_i=1|L^{j}_i)=\expit(\alpha_{j} L^{j}_i)$, where the subscript $i$ of $L^{j}_i$ denotes individual $i$ and $\expit(x) = \exp(x)/\{1+\exp(x)\}$.
Define the {\it inverse probability of treatment weight} \citep{rosenbaum1987model} for individual $i$ as:
\beq\label{eq:IPTW_def}
W_i^j = \dfrac{A_i}{\Pr\left(A_i=1|L_i=L^j_i\right)} + \dfrac{1-A_i}{1-\Pr\left(A_i=1|L_i=L^j_i\right)}.
\eeq
The weight $W_i^j$ is the reciprocal of the conditional probability of individual $i$ being assigned the observed treatment $A_i$ given the covariates $L^j_i$. 
Let $\widehat W_i^j$ denote the estimated weights obtained by substituting the maximum likelihood estimators (MLE) for $\alpha_{j}$ in the treatment model.
Fit the outcome regression model $\E(Y|A,L^{j}) = h^{-1}(\psi_{j}^\ast A + \beta_{j} L^{j})$ to the observed data using the aforementioned weights.
(The $\ast$ superscript indicates a conditional effect that may differ from the marginal effect $\psi_j$.)
Let $\widehat\E(Y|A,L^{j})$ denote the fitted outcome model obtained by plugging in the MLE for $\psi_{j}^\ast$ and $\beta_{j}$.
A doubly robust estimator of the average potential outcome $\E(Y^a)=n^{-1}\sum_{i} Y^a_i$ for $a=0,1$, is then:
\beq
\widehat\E(Y^a) =
n^{-1} \sum_{i} {\rm I}\{A_i=a\}\widehat W_i^j \left\{Y_i - \widehat\E(Y|A=A_i,L^{j}_i)\right\} + 
\widehat\E(Y|A=a,L^{j}_i),
\eeq
where ${\rm I}\{B\}=1$ if $B$ is true and $0$ otherwise.
The estimator for the marginal treatment effect $\psi=\E(Y^1)-\E(Y^0)$ in the $j$-th orbit is therefore:
\beq\label{eq:DRestimator_obs}
\widehat\psi_{j} = 
n^{-1} \sum_{i} (2A_i-1) \widehat W_i^j \left\{Y_i - \widehat\E(Y|A=A_i,L^{j}_i)\right\} + 
\widehat\E(Y|A=1,L^{j}_i) - \widehat\E(Y|A=0,L^{j}_i).
\eeq
When both treatment and outcome models are correctly specified, an asymptotic expansion around $\psi$ yields the so-called ``influence function'' for individual $i$ as:
\beq
\phi^{j}_i = (2A_i-1) W_i^j \left\{Y_i - \E(Y|A=A_i,L^{j}_i)\right\} + 
\E(Y|A=1,L^{j}_i)-\E(Y|A=0,L^{j}_i) - \psi.
\eeq
Let $\widehat\phi^{j}_i=(2A_i-1) \widehat W_i^j \left\{Y_i - \widehat\E(Y|A=A_i,L^{j}_i)\right\} + 
\widehat\E(Y|A=1,L^{j}_i) - \widehat\E(Y|A=0,L^{j}_i) - \widehat\psi_{j}$ denote the estimated influence function, obtained by plugging in the maximum likelihood estimators for the coefficients in the treatment and outcome models, and substituting the population expectation with a sample average.
The variance of the difference between effect estimators from two different orbits, e.g., $j$ and $k$, is consistently estimated by the sample variance (denoted by $\widehat\V$) of the corresponding difference in estimated influence functions:
\beq\label{eq:var_diff_infn}
\widehat\V\left\{n^{1/2}\left(\widehat\psi_{j}-\widehat\psi_{k}\right)\right\}
=
(n-1)^{-1} \sum_i \left(\widehat\phi^{j}_i - \widehat\phi^{k}_i\right)^2.
\eeq
Consistency and asymptotic normality of the standardized difference \eqref{eq:score_diff_std_fullmodel} with mean zero and variance one directly follow from the law of large numbers and the central limit theorem.
When both the treatment and outcome are continuous, and linear models for both may be reasonably assumed, the (doubly robust) ordinary least squares effect estimator, and its (asymptotic) variance estimator, are described in Appendix~\ref{sect:est_linear}.

\subsection{Randomization inference by full matching on the propensity score}\label{sect:reviewRI}

Following selection of the confounders, randomization inference for the causal effect can be readily carried out, using established methods for matching on the PS model\citep{stuart2010matching} given the selected covariates.
Under the randomization inference framework, the individual potential outcomes are assumed to be fixed (and possibly unknown) quantities, with treatment assignment being the only source of variability \citep{neyman1923}. 
Individuals with different treatments but otherwise similar PS (and therefore observed characteristics) are grouped together in the same stratum, thereby permitting repeated treatment assignments within each stratum toward valid inference about the causal effect. 
A key advantage of randomization inference is that it does not require invoking any assumptions of random sampling from some hypothetical (super)population and is therefore, appealing in observed study populations with no (known) well-defined sampling procedure.
In this paper, we will use {\it full matching}, which employs a specified propensity score model $\Pr(A_i=1|L_i)$ for a given set of covariates $L_i$ to create a collection of matched groups, where each matched group is guaranteed to contain at least one treated individual and at least one untreated individual. Each matched group can therefore be viewed as a single stratum \citep{fogarty2017randomization}. Full matching is optimal in terms of minimizing the average of the covariate distances (and across all covariates) between treated and untreated individuals within each matched group, and is particularly effective at reducing bias due to observed confounding variables \citep{austin2017estimating,austin2017performance}.
Full matching extends the concept of blocking \citep{Speed1992,morgan2012rerandomization} which uses the unique levels of a (few) covariate(s) to conduct completely randomized experiments within each block. 
Full matching therefore inherits the desirable properties of blocking, such as ensuring that there are any treatment assignments that do not preserve the observed covariate balance (in terms of matched propensity scores) have probability zero of occurring \citep{Rosenbaum:2002}.

Under the causal null of no individual treatment effect, defined as:
\beq
H_0: Y^1_i=Y^0_i, \, i=1,\ldots,n,
\eeq
the observed outcomes should be similarly distributed between treated and untreated individuals within each stratum. 
Let ${\bm A} = (A_1,\ldots,A_n)$ denote the treatment vector of length $n$, and let ${\bm a} = (a_1,\ldots,a_n)$ denote a possible value for ${\bm A}$. 
The plausibility of $H_0$ can therefore be assessed by calculating the frequency of obtaining a test statistic that is at least as `extreme' (from $H_0$) as its observed value assuming $H_0$, over hypothetical assignments of $\bm A$ holding the number assigned to treatment within each stratum fixed.
Denote the resulting set of hypothetical treatment assignments by $\Omega$, where each assignment occurs with equal probability $|\Omega|^{-1}$ under conditional exchangeability.
Let $X_i \in \{1,\ldots,R\}$ denote the stratum membership of individual $i$ when the observed sample is partitioned into $R$ strata. We adopt the (weighted) sum of the stratum-specific total outcomes for treated individuals as the test statistic; i.e., $\hat\tau({\bm A}) \!=\! n^{-1}\sum_r n_r \sum_i {\rm I}\{X_i \!=\! r\} A_i Y_i$, where $n_r$ is the total number of individuals in stratum $r$.
A (two-sided) p-value may then be defined as
$	{\rm pv}(\Omega) \!=\! |\Omega|^{\!-1}
        \underset{{\bm a} \in \Omega}{\sum}
        {\rm I}\left\{|\hat\tau({\bm a})| \!\geq\! |\hat\tau({\bm A})|\right\}$, 
where $\hat\tau({\bm a})$ is the test statistic obtained by replacing $A_i$ in $\hat\tau({\bm A})$ with $a_i$ under treatment $\bm a$, and larger (absolute) values of $\hat\tau({\bm A})$ suggest stronger evidence against $H_0$.
When it is not computationally feasible to enumerate $\Omega$ exactly - as is often in most realistic settings - an approximation based on e.g., $C=1000$ random draws of $\bm a$ from $\Omega$ may be used instead.

\section{Simulation studies}\label{sect:sims}
Simulation studies were conducted under different data-generating scenarios to empirically evaluate the ability of the proposal to (i) correctly select confounders, and (ii) yield hypothesis tests that preserved the type I error rate after data-driven covariate selection. 
Let $p$ denote the number of candidate observed baseline covariates. We partitioned the covariates into four subsets as follows.
Confounders in the first subset simultaneously affected treatment and outcome; their indices were denoted by ${\cal S}_1 = \{1,2\}$.
Covariates in the second subset affected outcome only; their indices were denoted by ${\cal S}_2 = \{3, 4\}$.
Instruments in the third subset were associated with treatment only; their indices were denoted by ${\cal S}_3 = \{5, 6\}$.
Covariates in the fourth subset were unassociated with either treatment or outcome; their indices were denoted by ${\cal S}_4 = \{7, \ldots, p\}$. The subsets were used exclusively to generate the observed data and not to inform any applied method.

We describe the study for one scenario and defer details of other scenarios.
Datasets with sample size $n=80$ and $p=25$ candidate covariates were generated under the null of no individual treatment effect, i.e., $H_0: Y^1_i=Y^0_i, \, i=1,\ldots,n$, as follows.
Let $L_{is}$ denote the $s$-th observed baseline covariate for individual $i$. For $s = 1,\ldots,p$, each covariate was randomly (and independently) draw from a standard normal distribution; i.e., $L_{is} \sim {\cal N}(0,1)$.
Denote the resulting vector of $p$ covariates by $L_i = (L_{i1},\ldots,L_{ip})$.
The (true) propensity score for each individual was determined as $\Pr(A_i=1|L_{i}) = \expit\left(\sum_{s=1}^{p} \gamma_s L_{is}\right)$, where $\gamma_s=1.0$ if $s \in {\cal S}_1$ (a confounder), or $\gamma_s=1.6$ if $s \in {\cal S}_3$ (an instrument), or $0$ otherwise. 
The observed treatment was then randomly drawn as $A_i \sim \hbox{Bernoulli}\{\Pr(A_i=1|L_{i})\}$.
The underlying outcome was determined as $Y_i^\ast = \sum_{s=1}^{p} \beta_s L_{is}$, where $\beta_s=0.8$ if $s \in {\cal S}_1 \cup {\cal S}_2$ (a confounder or an outcome-only predictor), or $0$ otherwise.
The observed outcome was then randomly drawn as $Y_i \sim {\cal N}(Y_i^\ast, 4^2)$.

We compared the ability of the proposal to correctly select the true confounders with the following variable selection methods that were implemented in publicly available software packages on the Comprehensive R Archive Network. We briefly describe each method and refer readers to the respective references for further details.
\ben[label=S\arabic*.]
\setcounter{enumi}{-1}
\item The stability-based confounder selection procedure proposed in this article (`Stability').

\item The Augmented Backward Elimination (`ABE') algorithm \citep{dunkler2014augmented} that combines backward variable selection with a change-in-estimate criterion. A significance-based threshold for a (standardized) regression coefficient of interest is used to determine the final model.
Because the method carries out variable selection for a single regression model, we used the outcome model where treatment was a `passive' variable that was always in the final model. 
The algorithm is implemented in the \texttt{abe} function\citep{blagus2017abe} (\url{https://CRAN.R-project.org/package=abe}). We set all options to their default levels.

\item The feature selection algorithm using random forests \citep{breiman2001random,rangerR2017} as implemented in the \texttt{Boruta} function \citep{kursa2010feature} (\url{https://CRAN.R-project.org/package=Boruta}). 
The method starts with the full model and iteratively compares the relative `importance' of observed variables with those of randomly generated non-informative variables, where the latter are created by permuting the former. Observed variables that have significantly worst importance than randomly generated ones are consecutively dropped. Observed variables that have significantly better importance are decided as `Confirmed.' Because the method carries out variable selection for a single regression model, we used the outcome model (starting with the full model conditional on treatment and all candidate covariates). Covariates that were decided as `Confirmed' were selected.

\item The covariate selection approach using the model-free backward elimination algorithms of De Luna et al\cite{de2011covariate} for nonparametric estimation of the marginal treatment effect. Covariates were selected based on whether there was significant evidence that they were in at least one of the minimal subsets ${\bf Q}_0$ and ${\bf Q}_1$ that rendered treatment conditionally independent of the potential outcomes $Y^0$ (in the control group) and $Y^1$ (in the treatment group)  respectively \citep{de2011covariate}. 
The algorithms are implemented in the \texttt{cov.sel} function from the \texttt{CovSel} package \citep{haggstrom2015covsel} (\url{https://CRAN.R-project.org/package=CovSel}). 
Because all the candidate covariates were continuous, we used marginal co-ordinate hypothesis tests (\texttt{type = "dr"}). 
Due to the small sample size, only algorithm 1 could be used \citep{haggstrom2019covsel}; we set all other options to their default levels, including the pre-determined significance level of $0.1$.  

\item The Consistent Significance Controlled Variable Selection algorithms\citep{zambom2018consistent} that select only significant variables in a linear regression model. Because different algorithms are available in the \texttt{SignifReg} function \citep{kim2019signifreg} (\url{https://CRAN.R-project.org/package=SignifReg}), we describe the algorithm corresponding to the default levels of the options. Forward selection starting with the intercept-only outcome model (because treatment could not be forced into the starting model) was used. At each step, the variable that generated a new model having the smallest maximum p-value among the Wald tests for the regression coefficients was added to the current model. This step was repeated so long as every regression coefficient was significant after correcting for multiple testing using the false discovery rate. 

\een

\begin{figure}[!ht]
    \centering
\includegraphics[width=.49\linewidth,keepaspectratio]{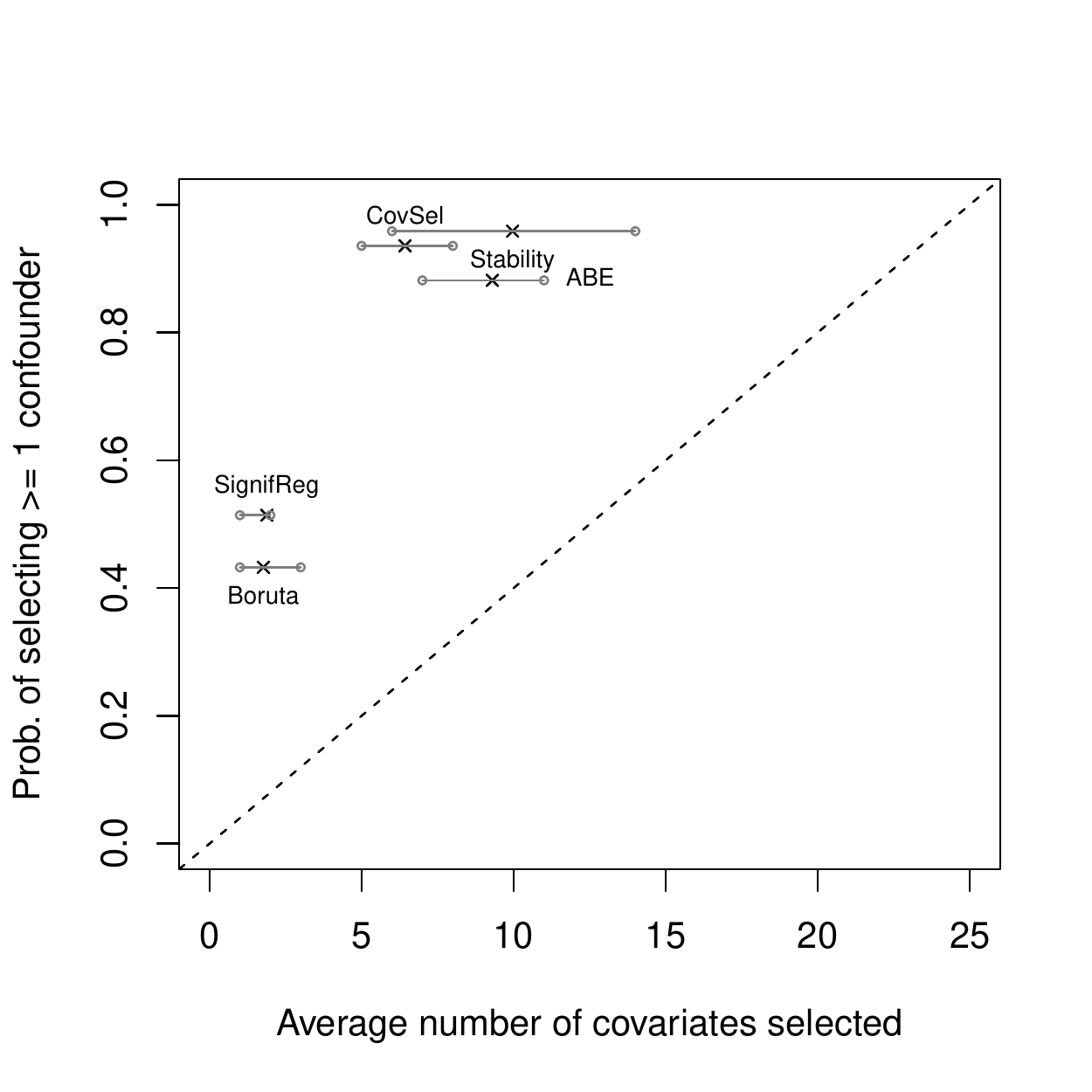}
\includegraphics[width=.49\linewidth,keepaspectratio]{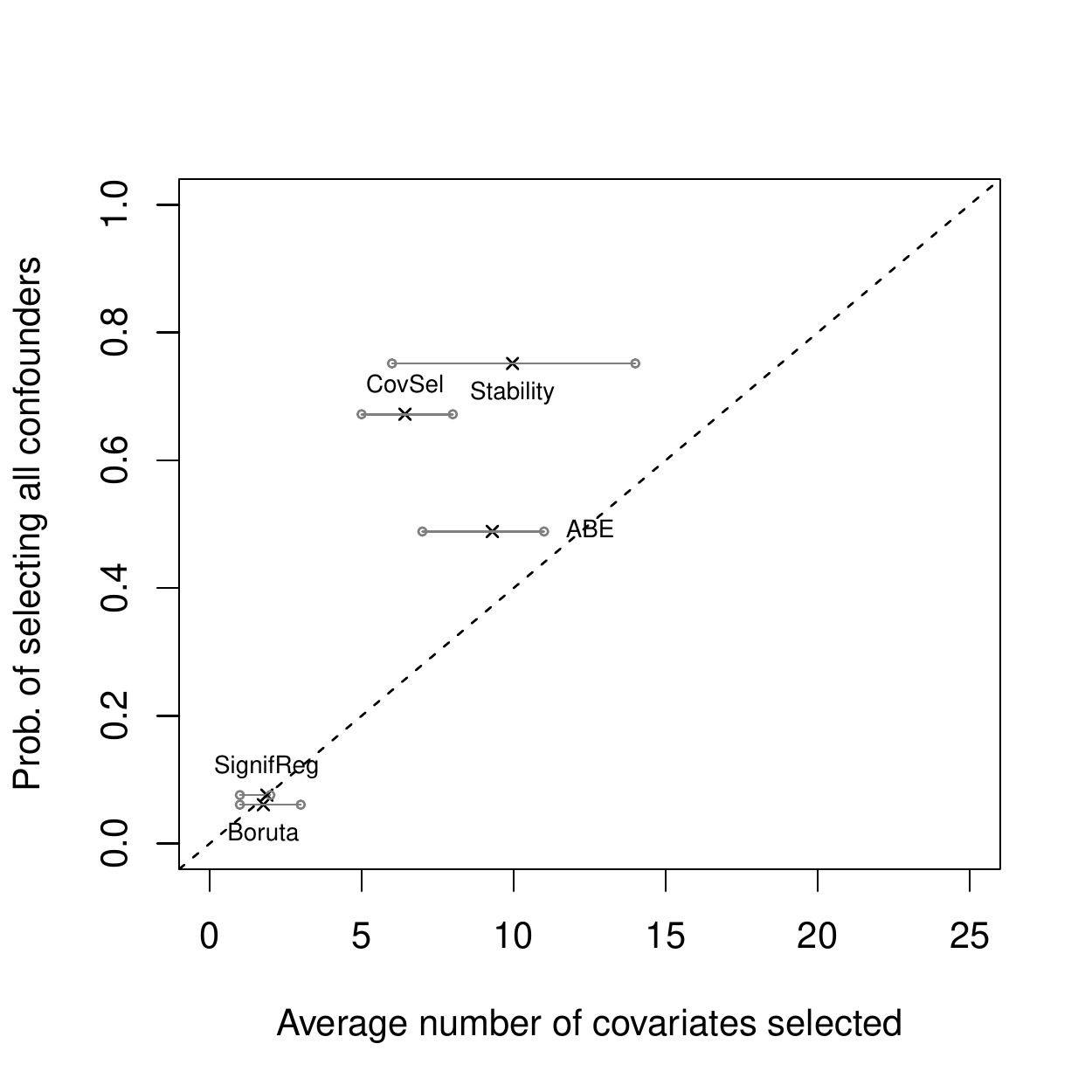}
    \caption{Scatterplot of the average empirical probabilities of selecting the true confounders (vertical axis) against the average number of selected covariates (horizontal axis), for each covariate selection method. The left and right panels respectively depict the proportions of simulated datasets where at least one confounder was included, and exactly both confounders were included, among the selected covariates. Each horizontal bar represents the interquartile range of the number of selected covariates.
    \label{fig:sims-selectVsPSsize}}
    \end{figure}

The average probabilities of including the (true) confounders among the selected covariates, with the average number of covariates selected, for each method are plotted in Figure~\ref{fig:sims-selectVsPSsize}. The Boruta and SignifReg methods selected the smallest number of covariates on average but failed to select any confounders about half the time. The ABE and CovSel methods selected at least one confounder almost as frequently as the proposed stability-based method, although ABE selected exactly both confounders only about half the time. 

We then assessed the ability of the selected covariates using each method to adjust for observed confounding, by including only (main effects for) the selected covariates in a PS model for full matching.
The resulting empirical distribution of the (randomization) p-values for testing $H_0$ (as described in Section~\ref{sect:reviewRI}) following the use of each data-driven covariate selection method was calculated. In addition, we considered the following methods for testing $H_0$:
\ben[label=M\arabic*.]
\item Full matching using the `target' PS model with only the true confounders ($L_{is}, s \in {\cal S}_1$) and the outcome-only predictors ($L_{is}, s \in {\cal S}_2$).
\item Randomization inference without adjusting for any covariates (`Empty PS model').
\item A Wald test using the sandwich-based standard errors for the treatment effect in a simple linear regression of outcome on treatment, with weights being the reciprocal of the fitted probabilities of the observed treatments from the outcome-adaptive LASSO procedure
\citep{shortreed2017outcome} (`OAL').
\item A Wald test using the (post-)double selection\cite{belloni2014inference} method for inference on (low-dimensional) target coefficients in a high-dimensional linear model. Penalized (linear) regression models for the outcome, and the treatment, on the observed covariates are first fitted, then the treatment effect is estimated by regressing the outcome on the treatment and all covariates with non-zero coefficient estimates in the fitted models.
This method (`HDM') is implemented in the \texttt{rlassoEffect} function of the \texttt{hdm} package \citep{chernozhukov2016hdm}.
\item A Wald test using discrete collaborative targeted minimum loss-based estimation (CTMLE) with greedy forward search \citep{van2010collaborative}. A (nested) sequence of candidate (parametric) PS models is constructed using a forward selection algorithm, by starting with the intercept-only model then adding (the main effect for) one additional covariate at each step. The targeted minimum loss estimator of the treatment effect corresponding to each PS model in the sequence is calculated. Cross-validation is then used to select the PS model that minimizes a specified loss function for the estimator (e.g., its MSE using the sum of squared residuals).
This procedure (`CTMLE: discrete') is implemented in the \texttt{ctmleDiscrete} function of the \texttt{ctmle} package \citep{ju2017ctmle}. We implemented all options as recommended in the online package vignette (\url{https://cran.r-project.org/web/packages/ctmle/vignettes/vignette.html}).
\item A Wald test using CTMLE with LASSO by fitting a penalized logistic regression model to estimate the PS model\citep{cheng2019collaborative}. Instead of a forward selection approach, the sequence of candidate PS models is now indexed by different values of the LASSO regularization penalty. As in the discrete CTMLE approach above, the PS model that minimizes the (cross-validated) loss function of the treatment effect estimator is chosen. This procedure is implemented in the \texttt{ctmleGlmnet} function of the \texttt{ctmle} package. 

\begin{figure}[!ht]
    \centering
\includegraphics[width=\linewidth,keepaspectratio]{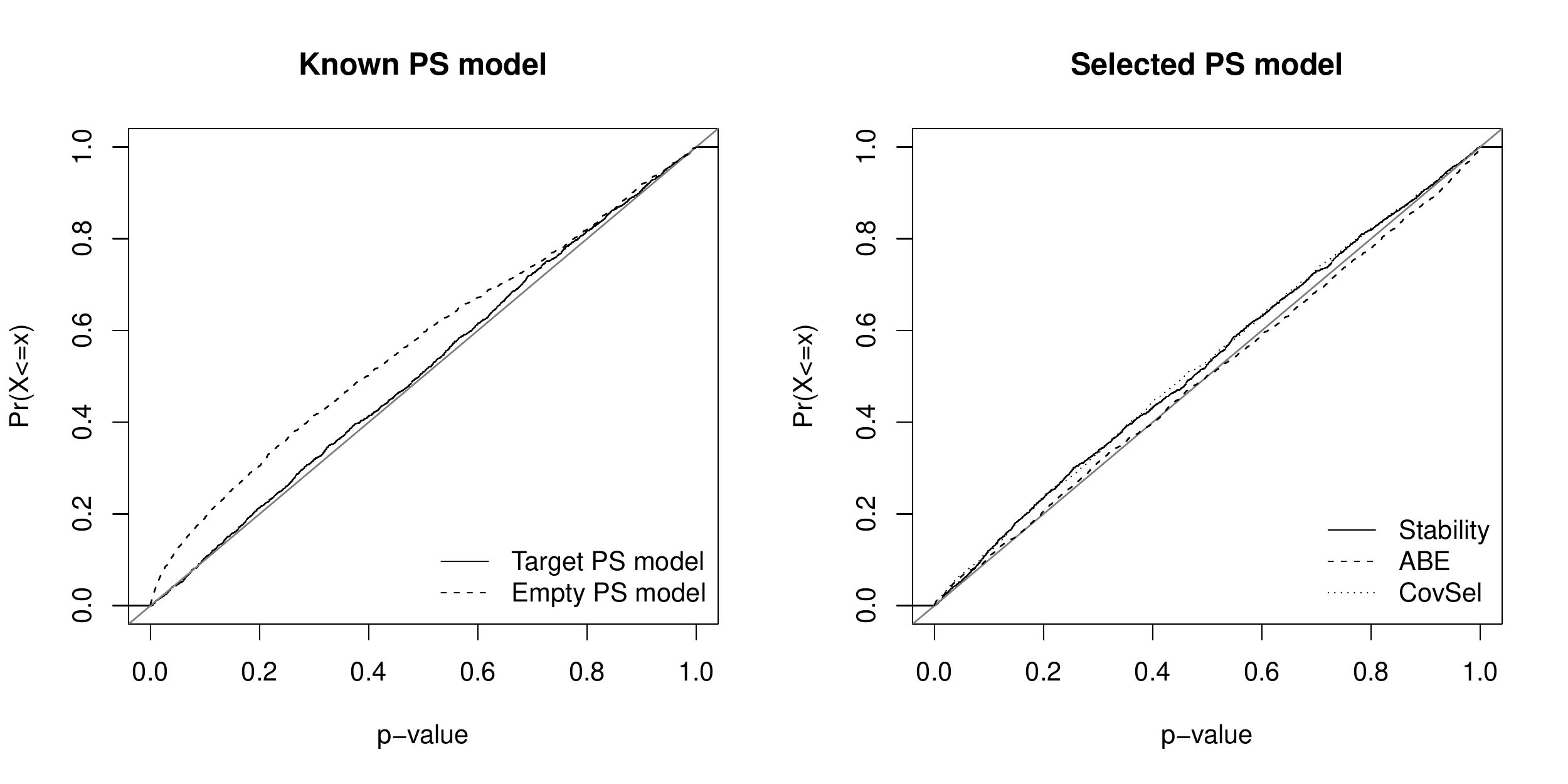}
    \caption{Empirical distributions of p-values for testing $H_0$. 
    The covariates in the PS model for full matching were either known (left panel), or selected using one of the plotted methods.
    The random variable $X$ on the vertical axis denotes the p-value.    
    The diagonal is plotted as a solid (gray) line with reduced opacity.\label{fig:sims-noselect}}
    \end{figure}

\een
Methods M1 and M2 did not require any covariate selection because the covariates in the PS model were either correctly specified (M1), or ignored (M2). The empirical distributions of the p-values for testing $H_0$ using these methods are plotted in the left panel of Figure~\ref{fig:sims-noselect}. 
As expected, failing to adjust for any covariates (M2) induced unobserved confounding of the treatment-outcome relation that resulted in inflated type I error rates; this is shown by the curve above the diagonal.
Randomization inference with full matching using the target PS model (M1) approximately controlled the type I error rate at its nominal level empirically; this is shown by the curve being close to the diagonal. 

\begin{figure}[!ht]
    \centering
\includegraphics[width=\linewidth,keepaspectratio]{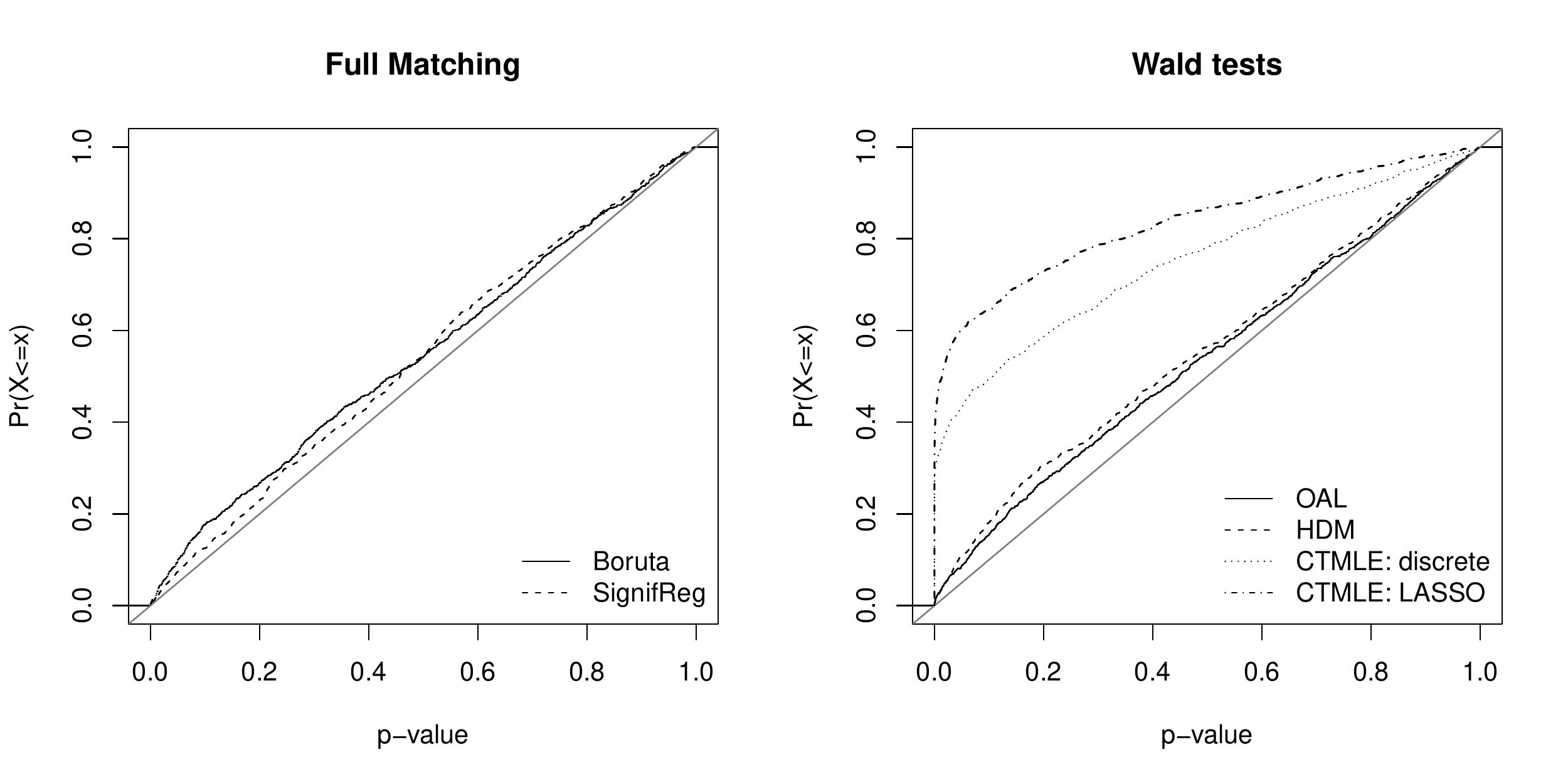}
    \caption{Empirical distributions of p-values for testing $H_0$. 
    In the left panel, different covariate selection methods were used to determine the PS (model) for full matching.
    In the right panel, inference for the treatment effect was carried out using either OAL, HDM, or one of the two versions of CTMLE.         
    The random variable $X$ on the vertical axes denote the p-value.
    The diagonal is plotted as a solid (gray) line with reduced opacity.
    \label{fig:sims-others1}}
    \end{figure}

The empirical distributions of the randomization p-values, following the use of the ABE, CovSel, and Stability covariate selection methods to determine the PS model for full matching, are plotted in the right panel of Figure~\ref{fig:sims-noselect}.
Because all three methods had similarly high probabilities of selecting (at least one of) the two true confounders, they approximately controlled the type I error empirically at all significance levels. 
In contrast, the Boruta and SignifReg methods resulted in inflated type I error rates as shown in the left panel of Figure~\ref{fig:sims-others1}.
Furthermore, none of the methods using (approximate) p-values from Wald tests of $H_0$ preserved the type I error rates empirically, as shown in the right panel of Figure~\ref{fig:sims-others1}.
The CTMLE methods also resulted in much smaller p-values than the HDM method on average.
To understand why, we compared the estimates of the average treatment effect and its standard error using these two approaches in Table~\ref{table:sims-ATE-compare}.
The standard error estimates using CTMLE were (much) smaller than those using HDM on average, possibly due in part to the `super-efficiency' properties of CTMLE which is aimed at optimizing the bias-variance tradeoff. However, the effect (point) estimates using CTMLE were more biased (and more variable) than those using HDM, resulting in larger empirical MSE (sum of the squared point estimate, and the variance estimate) on average. 
Hence the CTMLE effect estimates were seemingly more efficient whereby the Wald tests were rejected more frequently. 
For completeness, two Bayesian methods for treatment effect estimation following covariate selection were also considered: the Bayesian Adjustment for Confounding (BAC) algorithm as implemented in the \texttt{bac} function in the \texttt{bacr} package \citep{wang2015accounting}, and the Bayesian causal effect estimation algorithm \citep{talbot2015bayesian} as implemented in the \texttt{ABCEE} function in the \texttt{BCEE} package \citep{talbot2017bcee}.
Only the latter preserved the type I error rate empirically, but this is not guaranteed in general.
Details on their implementation, and the results, are provided in Appendix~\ref{sect:sim_results:appx}. 

\begin{table}[!ht]
\centering
\begin{tabular}{ cccccc } 
  Method & Mean effect estimate & E.s.e & Mean s.e. estimate & A.s.e & Mean MSE  \\
 \hline
 \hline 
 HDM & 0.59 & 1.29 & 1.20 & 0.18 & 1.76 \\ 
 CTMLE: discrete & 1.14 & 3.81 & 1.06 & 4.50 & 3.19 \\
 CTMLE: LASSO & 1.10 & 2.78 & 0.68 & 0.22 & 2.46 \\ 
 \hline 
 Stability & 0.16 & 1.78 & 0.70 & 0.25 & 1.67 \\ 
 ABE & 0.12 & 1.54 & 0.76 & 0.16 & 1.51 \\ 
 Boruta & 0.63 & 1.04 & 0.96 & 0.11 & 1.46 \\ 
 CovSel & 0.17 & 1.82 & 0.83 & 0.24 & 1.76 \\ 
 SignifReg & 0.40 & 0.99 & 0.90 & 0.10 & 1.30 \\
 \hline
\end{tabular}
\caption{Empirical summaries of estimates of the treatment effect and its standard error (s.e.) either using HDM (M5), or using CTMLE (M6), or following each of the confounder selection methods (S0 -- S4). The true value of the treatment effect was zero.
The empirical s.e. (E.s.e.) of the effect estimates, and asymptotic s.e. (A.s.e.) of the standard error estimates (i.e., its standard deviation across the simulated datasets), were considered. 
The average (square root of) the empirical MSE of the effect estimators were presented.
All results were rounded to two decimal places.
\label{table:sims-ATE-compare}}
\end{table}  

\subsection{Other settings}
The simulation study was extended to different settings with either (i) a binary outcome, where the observed outcome was randomly drawn as $Y_i \sim \hbox{Bernoulli}\{\expit(Y_i^\ast)\}$, or (ii) more covariates, e.g., $p=60$, or (iii) twice as many number of instruments, by setting ${\cal S}_3=(5,\ldots,8)$ and ${\cal S}_4=(9, \ldots, p)$.
In addition, we considered a different scenario by introducing certain covariates that induced ``collider bias\citep{pearl2009}'' when adjusted for. The data-generating mechanism was modified as follows, with all parameters adopting the same values as before unless otherwise stated.
Randomly draw two unobserved variables $U_{iv} \sim {\cal N}(0,1/16), v=1,2$.
For the observed covariates indexed by $s \in {\cal S}_3$, randomly draw $L_{is} \sim {\cal N}(2U_{i1}+2U_{i2},1/2)$, so that these covariates had (approximately) mean zero and unit variance. All other (observed) covariates were randomly drawn from the standard normal distribution as before.
The (true) propensity score for each individual was determined as $\Pr(A_i=1|L_{i}) = \expit\left(\sum_{s=1}^{p} \gamma_s L_{is} + \nu U_{i1}\right)$, where $\gamma_s=0, s \in {\cal S}_3$, and $\nu = 2$. 
The underlying outcome was determined as $Y_i^\ast = \beta_0 + \sum_{s=1}^{p} \beta_s L_{is} + \nu U_{i2}$. Hence the covariates $L_{is}$ indexed by $s \in {\cal S}_3$ were no longer instruments, but were now separately associated with treatment $A_i$, and outcome $Y_i$, through the unobserved common causes $U_{i1}$, and $U_{i2}$, respectively. Because these covariates were colliders along the path linking treatment and control, i.e., $A \leftarrow U_1 \rightarrow L_s \leftarrow U_2 \rightarrow Y$, controlling for these covariates would result in biased effect estimates. 

We summarize the findings here, and present the detailed results for each of the sixteen different data-generating scenarios in Appendix~\ref{sect:sim_results:appx}.
The proposed Stability method selected (exactly both) confounders more than 70\% of the time empirically for almost all settings, with the lowest probability of about 50\% when there were four instruments. Furthermore, when there were no instruments (and only covariates that may induce collider bias when adjusted for), the probability was more than 85\%. 
For almost all settings, the number of selected covariates was approximately half the total number of candidate covariates $p$, but when there were $p=60$ covariates and the outcome was continuous, the number of selected covariates was only about $p/4$ on average.
The Stability method therefore preserved the type I error rate empirically under almost all settings, except when there were $p=60$ covariates and the outcome was binary, due in part to overfitting in the logistic outcome model.

The ABE method was less likely to select (exactly both) confounders than the Stability method when the outcome was continuous and there were only $p=25$ covariates. When the outcome was binary, or there were $p=60$ covariates, or both, the ABE method tended to select almost all the available covariates on average, thus selecting the true confounders (much) more frequently than the Stability method. For these reasons, the ABE method seemingly preserved the type I error rate empirically under certain settings.
The CovSel method selected the confounders at a similar empirical probability as the Stability method when there were $p=25$ covariates, but at a much lower probability when there were $p=60$ covariates, which resulted in inflated type I error rates.
The Boruta and SignifReg methods managed to select (exactly both) confounders at best less than half of the time under certain settings, although the SignifReg method was seemingly able to preserve the type I error rate empirically under most settings but which may not hold in more general settings.
The CTMLE and HDM methods were considered only in settings with $p=25$ covariates, where their empirical type I error rates were inflated.

In conclusion, we demonstrated empirically that the proposed stability-based strategy either correctly included (both) confounders more frequently,  or selected a smaller subset of covariates, or both, than the other covariate selection methods under comparison. Furthermore, using full matching on the selected covariates empirically preserved the type I error rate approximately following data-driven covariate selection. 

\section{Illustration with applied examples}\label{sect:examples}

The proposed confounder selection strategy was illustrated using three publicly available datasets. 
For comparison, the covariates selected using each of ABE, Boruta, CovSel\footnote{
To allow for both continuous and non-continuous covariates in these data, kernel-based smoothing (\texttt{type = "np"}) with algorithm 2 was used for the CovSel model-free backward elimination method. Covariates were selected if there was significant evidence, at the (default) significance level of $0.1$, that they were in at least one of the minimal subsets ${\bf Z}_0$ and ${\bf Z}_1$ that rendered treatment conditionally independent of the potential outcomes $Y^0$ (in the control group) and $Y^1$ (in the treatment group) respectively \citep{de2011covariate}. All other options were set at their default values. 
}, 
and SignifReg methods were also determined.
Following each data-driven covariate selection method, full matching on the PS model containing only (main effects for) the selected covariates was used to construct strata. Randomization p-values for testing the null of no (individual) treatment effect were obtained using $C=2000$ assignments.
Because the `target' PS models were unknown in practice, the results using a (full) PS model containing all the available covariates were presented for comparison.
In addition, the (marginal) treatment effect estimator \eqref{eq:DRestimator_obs}, and its asymptotic variance estimator, given the selected covariates were calculated. 
However, we emphasize that (asymptotic) inference for the (standardized) treatment effect by merely using critical values from a standard normal distribution is unlikely to be valid, even after correctly selecting the covariates that adjust for all confounding, because the variance estimator does not reflect the uncertainty induced by carrying out covariate selection on the same data.
Nonetheless, we expect approximately valid inference by combining (i) double selection to prioritize covariates, (ii) focusing on the stability of the marginal effect estimator, and (iii) employing randomization inference.

\subsection{AIDS Clinical Trials Group Study 175}

The `ACTG175' dataset was from an AIDS randomized clinical trial, and was distributed as part of the \texttt{speff2trial} package via the Comprehensive R Archive Network (\url{https://CRAN.R-project.org/package=speff2trial}). The trial compared monotherapy using either zidovudine or didanosine alone with combination therapy using either zidovudine and didanosine, or zidovudine and zalcitabine, in adults infected with the human immunodeficiency virus type I whose CD4 T cell counts were between 200 and 500 per cubic millimeter. Treatment was (re)coded as $A=0$ for therapy using either zidovudine or didanosine only, and $A=1$ for therapies combining zidovudine and either didanosine or zalcitabine. A binary outcome was defined based on whether a participant's CD4 T cell count at $96\pm 5$ weeks was greater than 250 or not. The full dataset contained 2139 participants with 17 candidate (baseline) covariates, but we only considered a reduced dataset with 1342 participants having complete data so that a PS model with all covariates could be fitted. In addition, one covariate (prior zidovudine use) that was singular in the reduced dataset was dropped.

\begin{figure}[!ht]
    \centering
\includegraphics[width=.8\linewidth,keepaspectratio]{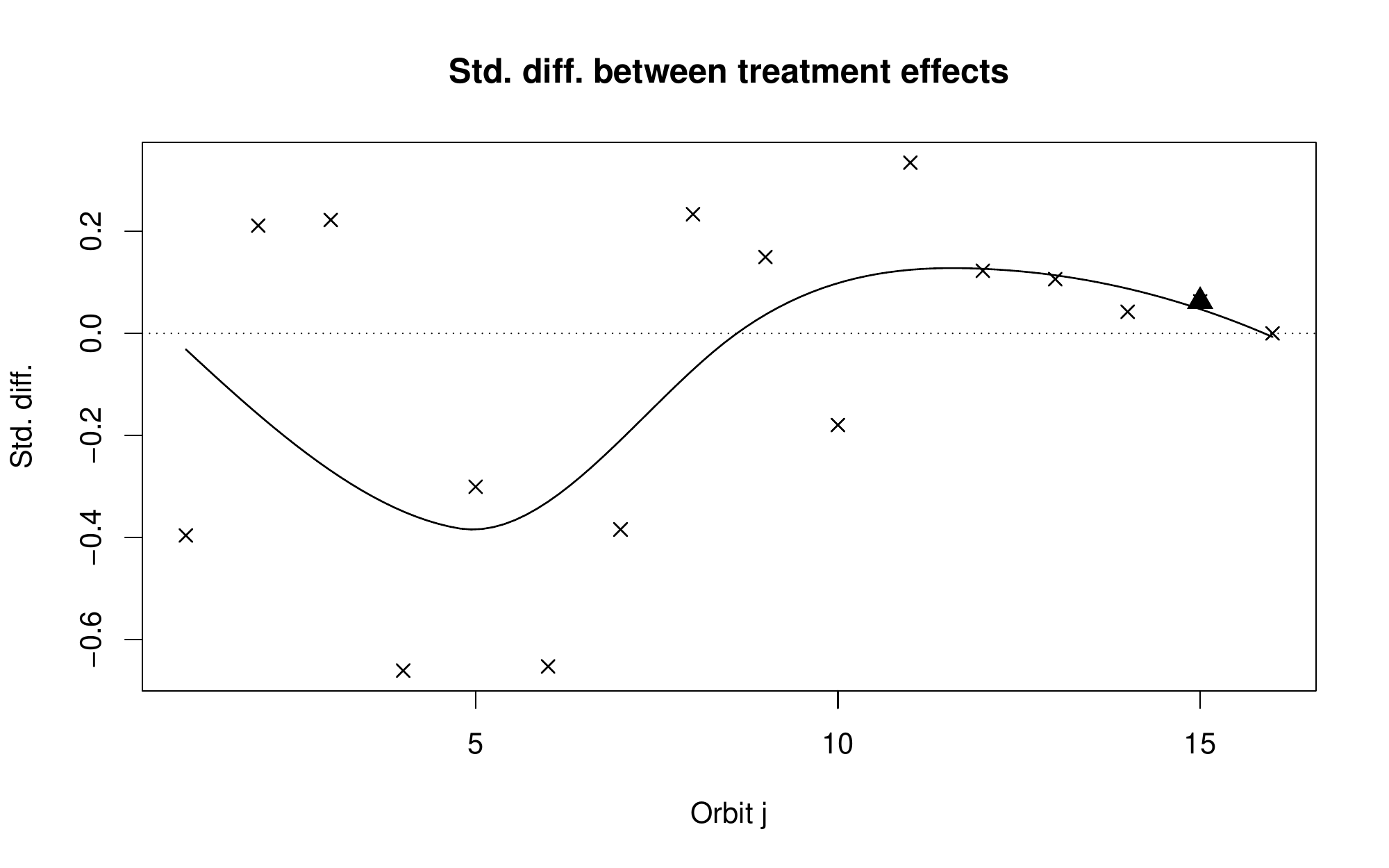}
    \caption{Standardized difference (`Std. diff.;' \eqref{eq:score_diff_std_fullmodel}), between the treatment effect estimator from each orbit and from the largest orbit, for the ACTG175 data. The solid black curve is a local cubic polynomial smoother. The most stable orbit minimizing Cochran's Q \eqref{eq:diagnostic_min_QJ} is indicated by a filled triangle. \label{fig:plot-speff2-score-vsAllL}}
    \end{figure}

The standardized differences \eqref{eq:score_diff_std_fullmodel} for each orbit (versus the largest orbit) are plotted in Figure~\ref{fig:plot-speff2-score-vsAllL}. 
The differences, which may be viewed as approximate biases for the estimators in each orbit, tended to fluctuate around zero for smaller orbits, before stabilizing (just above zero) in the larger orbits. Because this was a randomized controlled trial, confounding of the treatment-outcome relation was unlikely, but could have been induced by the exclusion of incomplete observations. Adjusting for different covariates therefore did not greatly affect the treatment effect estimates, as shown by the relatively small magnitudes of the differences (less than one standard deviation from zero).
The ordered covariates are displayed in Table~\ref{table:speff2-compare}. 
Because of the relatively few covariates, we considered windows of width three when calculating the Cochran's Q statistic in \eqref{eq:Q_diff_std_fullmodel} to evaluate stability. 
Although treatment was randomly assigned, the orbit that was deemed to be most stable contained 15 (out of 16 observed) covariates.
The results given the selected covariates using each of the methods, as well as without any covariates (`Empty'), are included in Table~\ref{table:speff2-compare}.
Across all of the covariate selection methods, the estimated effect was about 0.10, with the randomization p-value for testing $H_0$ much smaller than a significance level of e.g., 0.01, thus suggesting strong evidence that combination therapy had an effect on CD4 T cell count. 

\begin{table}[!ht]
\centering
\begin{tabular}{ |c|c|c|c|c|c|c|c| } 
 \hline
 Covariates & Stable & ABE & Boruta & CovSel & SignifReg & Empty & All\\\hline
CD4 T cell count at baseline (cd40) & TRUE & TRUE &  &  & TRUE &  & TRUE \\ 
antiretroviral history (str2) & TRUE & TRUE & TRUE &  & TRUE &  & TRUE \\ 
CD8 T cell count at baseline (cd80) & TRUE & TRUE &  &  & TRUE &  & TRUE \\ 
Karnofsky score (karnof) & TRUE & TRUE &  & TRUE & TRUE &  & TRUE \\ 
weight in kg at baseline (wtkg) & TRUE & TRUE &  &  &  &  & TRUE \\ 
history of intravenous drug use (drugs) & TRUE &  & TRUE & TRUE &  &  & TRUE \\ 
age & TRUE & TRUE &  &  &  &  & TRUE \\ 
symptomatic indicator (symptom) & TRUE & TRUE & TRUE &  &  &  & TRUE \\ 
history of antiretroviral therapy (preanti) & TRUE &  &  &  &  &  & TRUE \\ 
homosexual activity (homo)  & TRUE & TRUE &  &  &  &  & TRUE \\ 
gender & TRUE & TRUE & TRUE & TRUE &  &  & TRUE \\ 
race & TRUE &  &  & TRUE &  &  & TRUE \\ 
hemophilia (hemo) & TRUE &  &  &  &  &  & TRUE \\ 
prior non-zidovudine therapy (oprior) & TRUE &  & TRUE &  &  &  & TRUE \\ 
zidovudine use in last 30 days (z30) & TRUE &  & TRUE &  &  &  & TRUE \\ 
stratified antiretroviral history (strat) &  &  & TRUE &  &  &  & TRUE \\ 
\hline
 p-value for testing $H_0$ & $< 1 \!\times\! 10^{-3}$ & $2.5 \!\times\! 10^{-3}$ & $5 \!\times\! 10^{-3}$ & 
 $3 \!\times\! 10^{-3}$ & $< 1 \!\times\! 10^{-3}$ & $< 1 \!\times\! 10^{-3}$ & $< 1 \!\times\! 10^{-3}$ \\
 Treatment effect estimate & 0.104 & 0.106 & 0.104 & 0.090 & 0.103 & 0.096 & 0.104 \\
Standard error & 0.022 & 0.022 & 0.025 & 0.025 & 0.022 & 0.026 & 0.022 \\\hline
\end{tabular}
\caption{Comparison of the selected confounders (`TRUE') in the ACTG175 data using different methods. The covariates were ordered following double selection principles \eqref{eq:forwardselect_minimin}. 
The p-values for testing $H_0$, and estimated average (marginal) treatment effect and standard error (up to three decimal places), are stated in the last three rows. 
The results using a PS model with either no covariates (`Empty'), or all the available covariates (`All') are shown in the two rightmost columns.
\label{table:speff2-compare}}
\end{table}     

\subsection{LaLonde labor training program}

The `LaLonde' dataset was from a labor training program \citep{lalonde1986evaluating} used to demonstrate causal effect estimation adjusting for confounding; see e.g., Dehejia and Wahba\cite{dehejia1999causal} and Abadie and Imbens\cite{abadie2011bias}. 
We considered the version of the dataset that was distributed as part of the \texttt{CovSel} package \citep{haggstrom2015covsel}. There were 297 units assigned to participate in a national supported work demonstration ($A=1$), and 314 control units ($A=0$) drawn from survey datasets. The (continuous) outcome $Y$ was the level of post-intervention earnings in 1978. There were 10 candidate (baseline) covariates, with four continuous variables (age; years of schooling; and historical earnings in 1974 and 1975), and six binary variables (ethnicity being African-American, or Hispanic, or neither; martial status; high school diploma; and indicators of whether historical earnings in 1974 and 1975 were zero). 

\begin{figure}[!ht]
    \centering
\includegraphics[width=.8\linewidth,keepaspectratio]{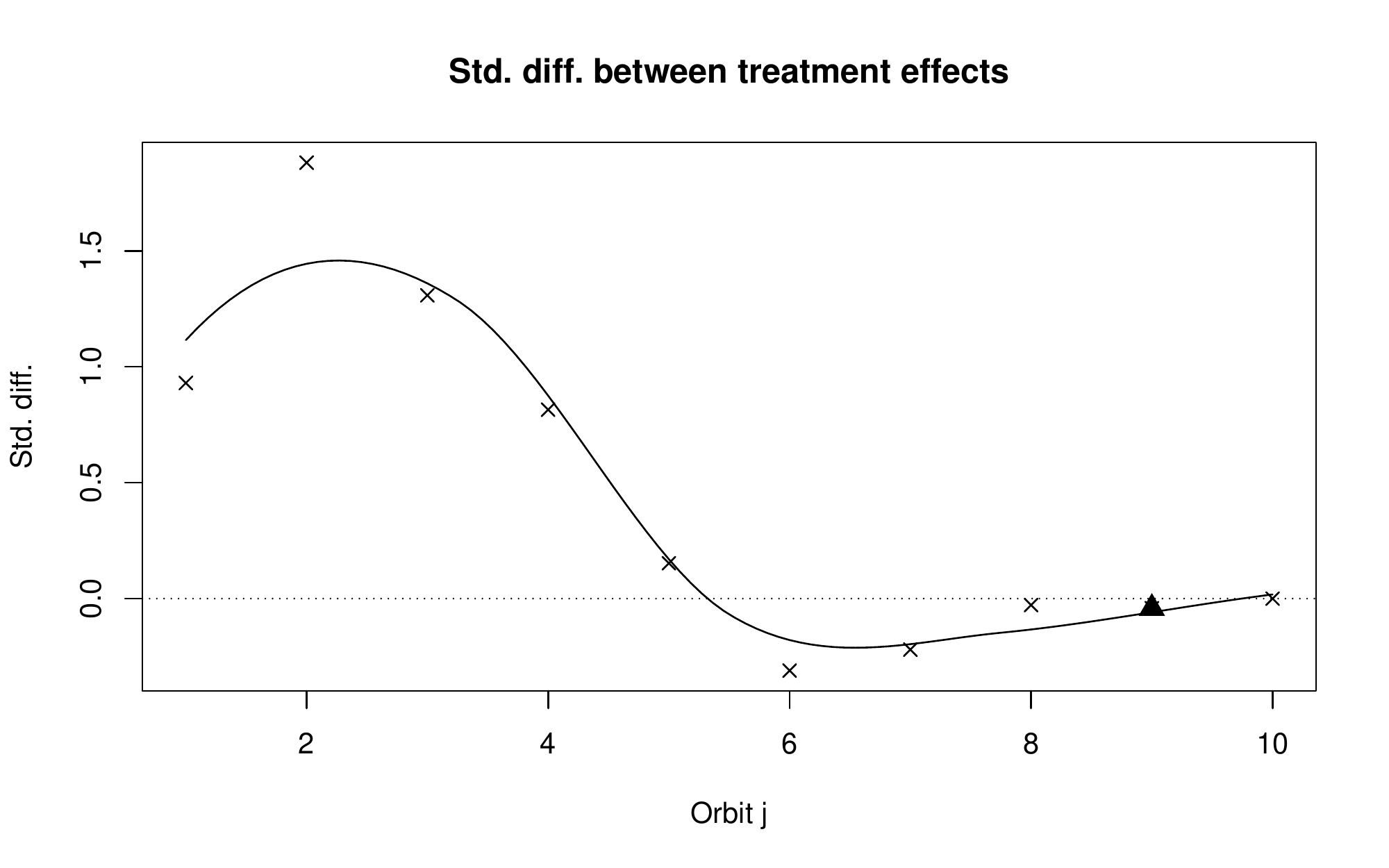}
    \caption{Standardized difference (`Std. diff.;' \eqref{eq:score_diff_std_fullmodel}), between the treatment effect estimator from each orbit and from the largest orbit, for the LaLonde data. The solid black curve is a local cubic polynomial smoother. The broken line indicates the 97.5\% quantile of the standard normal distribution.
    The most stable orbit minimizing Cochran's Q \eqref{eq:diagnostic_min_QJ} is indicated by a filled triangle. \label{fig:plot-lalonde-score-vsAllL}}
    \end{figure}

The standardized differences \eqref{eq:score_diff_std_fullmodel} for each orbit (versus the largest orbit) are plotted in Figure~\ref{fig:plot-lalonde-score-vsAllL}. 
There were no sudden changes in the differences between consecutive orbits, which tended to zero in the larger orbits. 
The ordered covariates are displayed in Table~\ref{table:lalonde-compare}.
Because of the relatively few covariates, we considered windows of width three when calculating the Cochran's Q statistic in \eqref{eq:Q_diff_std_fullmodel} to evaluate stability. 
The most stable orbit contained 9 (out of 10 observed) covariates.
The results given the selected covariates using each of the methods were included for comparison in Table~\ref{table:lalonde-compare}.
The randomization-based p-value for testing $H_0$ using the proposed stability(-based) strategy was far above reasonable significance levels, and in line with the results following the other methods, therefore suggesting little evidence that the work demonstration affected later earnings. 

\begin{table}[!ht]
\centering
\begin{tabular}{ |c|c|c|c|c|c|c| } 
 \hline
 Covariates & Stable & ABE & Boruta & CovSel & SignifReg & All \\\hline
 African-American ethnicity & TRUE & TRUE &  & TRUE &  & TRUE \\ 
 real earnings in 1974 & TRUE & TRUE & TRUE &  & TRUE & TRUE \\ 
 zero earnings in 1974 & TRUE & TRUE & TRUE & TRUE & TRUE & TRUE \\ 
 years of schooling & TRUE & TRUE &  & TRUE & TRUE & TRUE \\ 
 real earnings in 1975 & TRUE & TRUE & TRUE &  & TRUE & TRUE \\ 
 martial status & TRUE &  & TRUE & TRUE &  & TRUE \\ 
 Hispanic ethnicity & TRUE &  & TRUE &  &  & TRUE \\ 
 high school diploma & TRUE &  & TRUE & TRUE &  & TRUE \\ 
 zero earnings in 1975 & TRUE &  & TRUE & TRUE &  & TRUE \\ 
 age &  &  & TRUE & TRUE &  & TRUE \\ 
 \hline 
p-value for testing $H_0$ & 0.630 & 0.689 & 0.935 & 0.535 & 0.644 & 0.657 \\
Treatment effect estimate & -182 & -114 & -669  & 154 & -900 & -182 \\
Standard error & 881 & 776 & 715 & 766 & 675 & 886 \\ \hline
\end{tabular}
\caption{Comparison of the selected confounders (`TRUE') in the LaLonde data using different methods. The covariates were ordered following double selection principles \eqref{eq:forwardselect_minimin}. 
The p-values for testing $H_0$ (up to three decimal places), and estimated average (marginal) treatment effect and standard error (nearest integer), are stated in the last three rows. 
The results using a PS model with all the available covariates are shown in the rightmost column (`All').
\label{table:lalonde-compare}}
\end{table}

\clearpage

\subsection{Right Heart Catheterization}

The `RHC' dataset was from an observational study on the effectiveness of Right Heart Catheterization (RHC) in the initial care of critically ill patients \citep{connors1996effectiveness}, and was distributed as part of the \texttt{Hmisc} package via the Comprehensive R Archive Network (\url{https://CRAN.R-project.org/package=Hmisc}).
The data contained information on hospitalized adult patients at five medical centers in the U.S. who participated in the Study to Understand Prognoses and Preferences for Outcomes and Risks of Treatments (SUPPORT). The treatment variable was defined to be whether or not a patient received an RHC within 24 hours of admission. A binary outcome was defined based on whether a patient died at any time up to 180 days since admission. The full dataset contained 5735 participants with 73 covariates. However, to fit the PS model with all available covariates, we considered a reduced dataset with 2707 participants having complete data on 72 covariates (one covariate that was singular in the reduced dataset was dropped). 

\begin{figure}[!ht]
    \centering
\includegraphics[width=.8\linewidth,keepaspectratio]{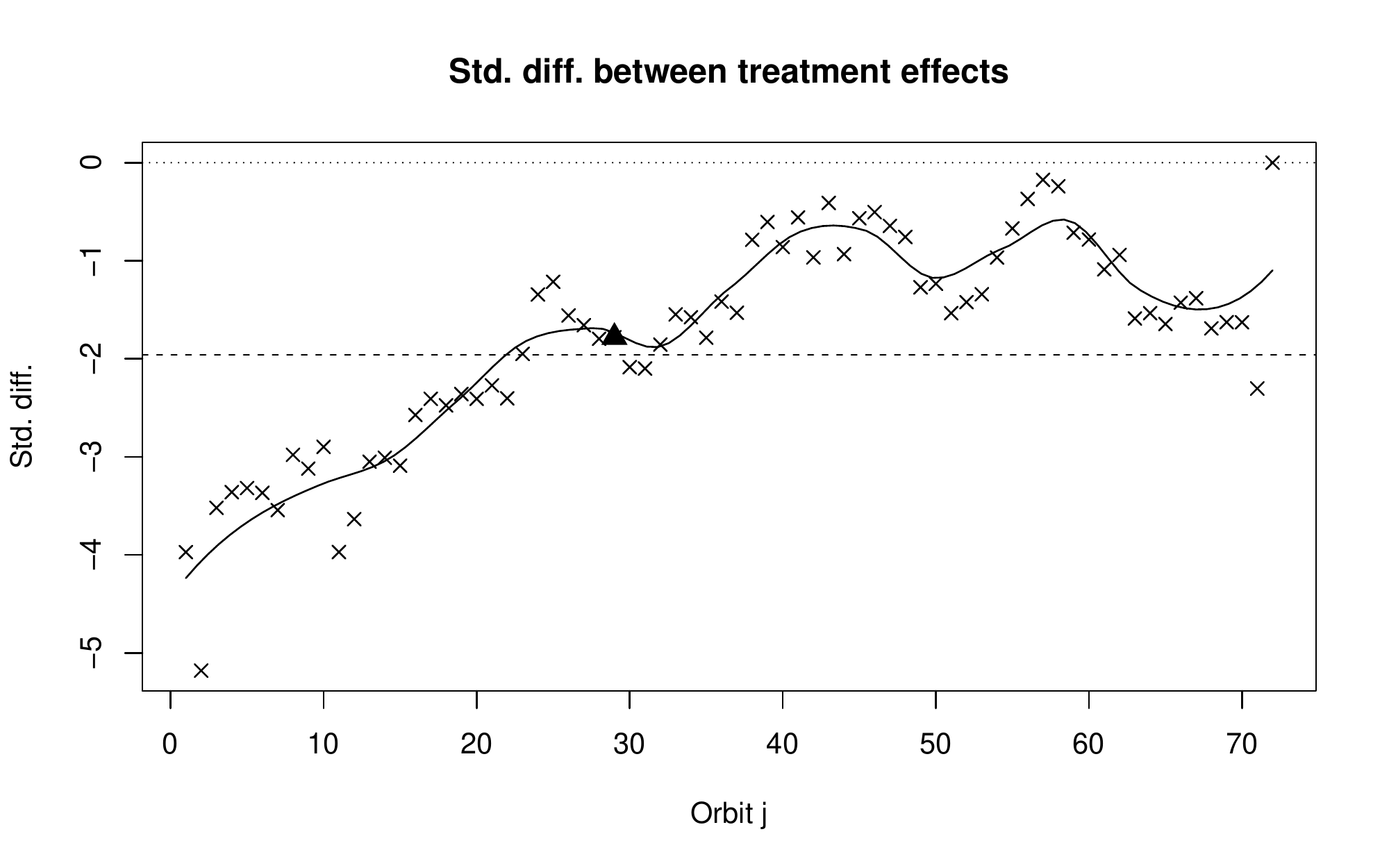}
    \caption{Standardized difference (`Std. diff.;' \eqref{eq:score_diff_std_fullmodel}), between the treatment effect estimator from each orbit and from the largest orbit, for the RHC data. The solid black curve is a local cubic polynomial smoother. The broken line indicates the 2.5\% quantile of the standard normal distribution.
    The most stable orbit minimizing Cochran's Q \eqref{eq:diagnostic_min_QJ} is indicated by a filled triangle. \label{fig:plot-rhc-score-vsAllL}}
    \end{figure}

\begin{figure}[!ht]
    \centering
\includegraphics[width=.8\linewidth,keepaspectratio]{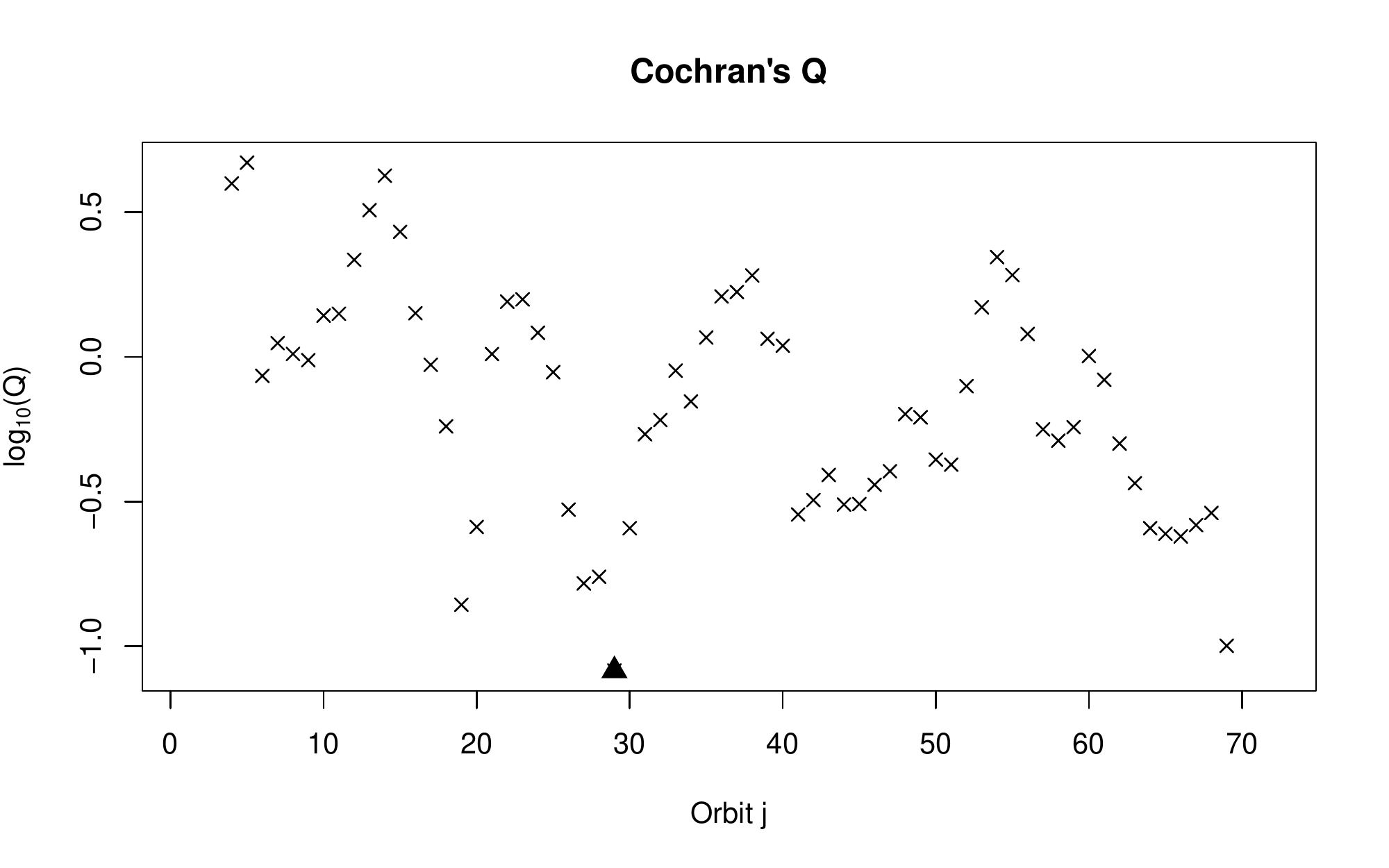}
    \caption{Values of Cochran's Q \eqref{eq:Q_diff_std_fullmodel} for each orbit $j$ for the RHC data. The minimal value is indicated by a filled triangle. \label{fig:plot-rhc-score-vsAllL-2-minimin-Q}}
    \end{figure}

The standardized differences \eqref{eq:score_diff_std_fullmodel} for each orbit (versus the largest orbit) are plotted in Figure~\ref{fig:plot-rhc-score-vsAllL}. 
It can be seen that the trajectory of the differences oscillated across consecutive orbits, and was relatively stable in different regions where the smoother could be considered to be relatively flat. 
Visually inspecting the fluctuations led us to consider windows of width seven when calculating the Cochran's Q statistic in \eqref{eq:Q_diff_std_fullmodel} to evaluate stability. The orbit deemed to be most stable contained 29 (out of 72 observed) covariates. 
The values of Cochran's Q plotted in Figure~\ref{fig:plot-rhc-score-vsAllL-2-minimin-Q} indicated that a larger orbit (with 69 covariates) was similarly stable, with a (locally minimal) value close to the (globally) minimal value attained by a smaller orbit (with 29 covariates).
When wider windows of width nine were used to evaluate stability, the selected orbit contained 28 covariates (one fewer than with windows of width seven).
For comparison, 34, 20 and 13 covariates were selected using the ABE, Boruta, and SignifReg methods respectively.
Due to space constraints, the ordered covariates, as well as those selected using each method, are displayed in Table~\ref{table:rhc-compare} in the Appendix. The CovSel method was omitted as it could not be completed successfully. 

Adjusting for the selected covariates using the proposed stability(-based) strategy yielded an estimated (marginal) treatment effect of 0.06, with a 95\% Wald confidence interval of $(0.016, 0.103)$ that excluded zero and suggested that RHC significantly affected mortality within 180 days. 
These results were similar to the findings using the matching and machine learning methods under comparison in Table 1 of Keele and Small\cite{keele2018comparing}.
However, full matching using a PS model with (only main effects for) the selected covariates yielded 664 strata, with  half containing just two individuals within each stratum. The thin strata limited the number of (unique) hypothetical treatment assignments, which resulted in a more conservative randomization p-value of 0.193 for testing $H_0$.

\section{Discussion}\label{sect:discuss}

In this article, we have proposed a confounder selection strategy that explicitly assesses the stability of the treatment effect estimator across different (nested) covariate subsets as a selection criterion.
The proposal was demonstrated empirically to yield approximately valid inference following a data-driven selection of covariates through the combined use of (i) double selection for prioritizing the covariates, (ii) stability-based assessment to select covariates for confounding adjustment, and (iii) randomization inference using full matching to control the type I error when testing the null of no (individual) treatment effect.

There are several avenues of possible future research related to the confounder selection strategy that focuses on the stability of the treatment effect estimator developed in this article.
When ordering the orbits in the first part, any covariate selection method that permits prioritizing the covariates for confounder selection can conceptually be incorporated into the proposed stability-based strategy.
For example, using double selection principles may reduce the chances of selecting predictors of outcome (only), because adjusting for such covariates will only improve the variance but not the bias of the treatment effect estimator in large samples.
To overcome this potential loss in efficiency, a different method that prioritizes such covariates that can curtail variance inflation, such as the outcome-adaptive LASSO, may be considered for ordering the covariates instead. For example, the covariates (standardized to have mean zero and variance one) may be ordered based on the absolute magnitude of their estimated (standardized) coefficients in the penalized PS model. Notwithstanding such possible improvements, we elected not to use this approach as we conjecture that it would unlikely deliver approximately valid inference as compared to double selection. 
Other approaches using the Bayesian framework may also be considered. For example, Wilson and Reich\cite{wilson2014confounder} propose a decision-theoretic approach to covariate selection via penalized credible regions of the regression parameters that form sets of (nested) candidate models. The solution paths based on the posterior probabilities of each covariate being selected may then be used to order the covariates. Similarly, Bayesian variable selection methods\citep{talbot2015bayesian}, and Bayesian model averaging methods\citep{wang2015accounting}, may also be used to order the covariates based on their decreasing posterior probabilities of inclusion in the PS model. 
Because different covariate selection methods employ different selection mechanisms that target different adjustment sets, comparing the abilities of different approaches to prioritize the covariates in delivering (approximately) valid inference is an area of future work.

In ongoing work, we show that one can interestingly gain insight into the sensitivity of the obtained results to unmeasured confounding, by extrapolating the fitted trajectory of the effect estimator across orbits, toward hypothetical orbit sizes larger than the size $J$ that was observed in the study.
Evaluating different methods or criteria for judging stability in the second part of the proposed procedure is another direction for future research. One possibility is to impose either pre-defined or data-adaptive thresholds on the values of the chosen statistic to limit the candidate orbits for consideration.
For example, in the illustration using the RHC data, the most stable orbit (see Figure~\ref{fig:plot-rhc-score-vsAllL}) was determined to be (just) inside the 95\% quantiles of the (asymptotically standard normally-distributed) standardized differences. One may consider evaluating stability only among orbits with standardized differences that are closer to zero, e.g., within one standard deviation.
The Q statistic as proposed in \eqref{eq:Q_diff_std_fullmodel} considered windows of width five, so that if the approximate bias $\widehat \psi_{k}-\widehat \psi_{J}$ is constant for $k=j-2,\cdots,j+2$, then even if each value of $\widehat \psi_{k}$ is far from $\widehat \psi_{J}$, the Q statistic will flag orbit $j$ as having a stable (but biased) effect estimator.
Other possibilities for judging stability include considering windows of different widths following visual inspection of the trajectory, as we carried out in the applied examples, or alternative criteria that measure stability differently, or directly optimizing the trajectory of the effect estimator.
For example, a local polynomial\citep{cleveland1992statistical}, or smoothing spline\citep{hastie1990generalized}, regression may be fitted to the effect estimates as a function of the orbits, and the critical points of the smoothing function can be evaluated numerically using finite difference approximations for the derivatives.
We acknowledge that default settings were simply employed for the other methods (for selecting confounders and post-selection effect estimation) under comparison in the simulation studies. These existing methods should be fine-tuned to improve their (simulation-based) performance in future neutral comparison studies \citep{boulesteix2018necessity}.

Finally, we have focused on testing the null of no (individual) treatment effect. In general, given a set of (selected) covariates in the PS model, randomization-based confidence intervals can be constructed by inverting a series of hypothesis tests under maintained hypotheses of assumed constant non-zero individual treatment effects. However, such an approach merely conditions on the (selected) covariates, and does not account for the uncertainty induced by the (data-driven) covariate selection procedure. In general, existing post-selection inference procedures (see, e.g., Berk et al.\citep{berk2013valid}, Chernozhukov et al.\citep{chernozhukov2015valid}, and Ju et al.\citep{cheng2019collaborative}, among others) typically consider regularized regression methods, and share the limitation that they lack finite sample guarantees and attain their desired theoretical properties only for large samples sizes.
Various aspects of our proposal (i.e., using double selection to prioritize covariates, focusing on the stability of the marginal effect estimator, and employing randomization inference) were chosen so that standard inference would deliver reasonable approximations in finite samples. While the validity of our proposal was confirmed in empirical studies, the complex nature of the problem makes it unlikely that this can be theoretically guaranteed in a uniform sense. 
Building on the results in this article, a good understanding of our proposal will therefore be enhanced by additional, extensive numerical studies under a diversity of data-generating mechanisms.

\section*{Acknowledgements}
The authors would like to thank the Editor, Associate Editor, and two reviewers for their comments on prior versions of this manuscript. 
This research was supported by the Special Research Fund (BOF) of Ghent University research project BOF.24Y.2017.0004.01.
Computational resources and services were provided by the VSC (Flemish Supercomputer Center), funded by the FWO and the Flemish Government -- department EWI.
The content is solely the responsibility of the authors and does not 
represent the official views of the authors' institutions.
\section*{Supporting information}

The \texttt{R} code used to implement the proposed procedure and other methods in carrying out the simulation studies in Section~\ref{sect:sims}, and in conducting the analyses in Section~\ref{sect:examples}, are available at the following web address: \url{https://github.com/wwloh/stability-confounder-select}

\bibliographystyle{abbrvnat}
\bibliography{../../ri_varselect}

\begin{thebibliography}{57}
\providecommand{\natexlab}[1]{#1}
\providecommand{\url}[1]{\texttt{#1}}
\expandafter\ifx\csname urlstyle\endcsname\relax
  \providecommand{\doi}[1]{doi: #1}\else
  \providecommand{\doi}{doi: \begingroup \urlstyle{rm}\Url}\fi

\bibitem[Abadie and Imbens(2011)]{abadie2011bias}
A.~Abadie and G.~W. Imbens.
\newblock Bias-corrected matching estimators for average treatment effects.
\newblock \emph{Journal of Business \& Economic Statistics}, 29\penalty0
  (1):\penalty0 1--11, 2011.

\bibitem[Austin and Stuart(2017{\natexlab{a}})]{austin2017estimating}
P.~C. Austin and E.~A. Stuart.
\newblock Estimating the effect of treatment on binary outcomes using full
  matching on the propensity score.
\newblock \emph{Statistical Methods in Medical Research}, 26\penalty0
  (6):\penalty0 2505--2525, 2017{\natexlab{a}}.

\bibitem[Austin and Stuart(2017{\natexlab{b}})]{austin2017performance}
P.~C. Austin and E.~A. Stuart.
\newblock The performance of inverse probability of treatment weighting and
  full matching on the propensity score in the presence of model
  misspecification when estimating the effect of treatment on survival
  outcomes.
\newblock \emph{Statistical Methods in Medical Research}, 26\penalty0
  (4):\penalty0 1654--1670, 2017{\natexlab{b}}.

\bibitem[Belloni et~al.(2014)Belloni, Chernozhukov, and
  Hansen]{belloni2014inference}
A.~Belloni, V.~Chernozhukov, and C.~Hansen.
\newblock Inference on treatment effects after selection among high-dimensional
  controls.
\newblock \emph{The Review of Economic Studies}, 81\penalty0 (2):\penalty0
  608--650, 2014.

\bibitem[Berk et~al.(2013)Berk, Brown, Buja, Zhang, Zhao,
  et~al.]{berk2013valid}
R.~Berk, L.~Brown, A.~Buja, K.~Zhang, L.~Zhao, et~al.
\newblock Valid post-selection inference.
\newblock \emph{The Annals of Statistics}, 41\penalty0 (2):\penalty0 802--837,
  2013.

\bibitem[Blagus(2017)]{blagus2017abe}
R.~Blagus.
\newblock \emph{abe: Augmented Backward Elimination}, 2017.
\newblock URL \url{https://CRAN.R-project.org/package=abe}.
\newblock R package version 3.0.1.

\bibitem[Boulesteix et~al.(2018)Boulesteix, Binder, Abrahamowicz, Sauerbrei,
  and for the Simulation Panel of~the
  STRATOS~Initiative]{boulesteix2018necessity}
A.-L. Boulesteix, H.~Binder, M.~Abrahamowicz, W.~Sauerbrei, and for the
  Simulation Panel of~the STRATOS~Initiative.
\newblock On the necessity and design of studies comparing statistical methods.
\newblock \emph{Biometrical Journal}, 60\penalty0 (1):\penalty0 216--218, 2018.
\newblock \doi{10.1002/bimj.201700129}.
\newblock URL
  \url{https://onlinelibrary.wiley.com/doi/abs/10.1002/bimj.201700129}.

\bibitem[Breiman(2001)]{breiman2001random}
L.~Breiman.
\newblock Random forests.
\newblock \emph{Machine Learning}, 45\penalty0 (1):\penalty0 5--32, 2001.

\bibitem[Brookhart et~al.(2006)Brookhart, Schneeweiss, Rothman, Glynn, Avorn,
  and St{\"u}rmer]{brookhart2006variable}
M.~A. Brookhart, S.~Schneeweiss, K.~J. Rothman, R.~J. Glynn, J.~Avorn, and
  T.~St{\"u}rmer.
\newblock Variable selection for propensity score models.
\newblock \emph{American Journal of Epidemiology}, 163\penalty0 (12):\penalty0
  1149--1156, 2006.

\bibitem[Cheng et~al.(2017)Cheng, Gruber, and {van der Laan}]{ju2017ctmle}
J.~Cheng, S.~Gruber, and M.~J. {van der Laan}.
\newblock \emph{ctmle: Collaborative Targeted Maximum Likelihood Estimation},
  2017.
\newblock URL \url{https://CRAN.R-project.org/package=ctmle}.
\newblock R package version 0.1.1.

\bibitem[Chernozhukov et~al.(2015)Chernozhukov, Hansen, and
  Spindler]{chernozhukov2015valid}
V.~Chernozhukov, C.~Hansen, and M.~Spindler.
\newblock Valid post-selection and post-regularization inference: An
  elementary, general approach.
\newblock 2015.

\bibitem[Chernozhukov et~al.(2016)Chernozhukov, Hansen, and
  Spindler]{chernozhukov2016hdm}
V.~Chernozhukov, C.~Hansen, and M.~Spindler.
\newblock {hdm}: High-dimensional metrics.
\newblock \emph{R Journal}, 8\penalty0 (2):\penalty0 185--199, 2016.
\newblock URL
  \url{https://journal.r-project.org/archive/2016/RJ-2016-040/index.html}.

\bibitem[Cleveland et~al.(1992)Cleveland, Grosse, Shyu, Chambers, and
  Hastie]{cleveland1992statistical}
W.~S. Cleveland, E.~Grosse, W.~M. Shyu, J.~M. Chambers, and T.~J. Hastie.
\newblock Statistical models in s.
\newblock \emph{Local regression models}, pages Chapter--8, 1992.

\bibitem[Connors et~al.(1996)Connors, Speroff, Dawson, Thomas, Harrell, Wagner,
  Desbiens, Goldman, Wu, Califf, et~al.]{connors1996effectiveness}
A.~F. Connors, T.~Speroff, N.~V. Dawson, C.~Thomas, F.~E. Harrell, D.~Wagner,
  N.~Desbiens, L.~Goldman, A.~W. Wu, R.~M. Califf, et~al.
\newblock The effectiveness of right heart catheterization in the initial care
  of critically iii patients.
\newblock \emph{Journal of the American Medical Association}, 276\penalty0
  (11):\penalty0 889--897, 1996.

\bibitem[Crainiceanu et~al.(2008)Crainiceanu, Dominici, and
  Parmigiani]{crainiceanu2008adjustment}
C.~M. Crainiceanu, F.~Dominici, and G.~Parmigiani.
\newblock Adjustment uncertainty in effect estimation.
\newblock \emph{Biometrika}, 95\penalty0 (3):\penalty0 635--651, 2008.

\bibitem[De~Luna et~al.(2011)De~Luna, Waernbaum, and
  Richardson]{de2011covariate}
X.~De~Luna, I.~Waernbaum, and T.~S. Richardson.
\newblock Covariate selection for the nonparametric estimation of an average
  treatment effect.
\newblock \emph{Biometrika}, 98\penalty0 (4):\penalty0 861--875, 2011.

\bibitem[Dehejia and Wahba(1999)]{dehejia1999causal}
R.~H. Dehejia and S.~Wahba.
\newblock Causal effects in nonexperimental studies: Reevaluating the
  evaluation of training programs.
\newblock \emph{Journal of the American Statistical Association}, 94\penalty0
  (448):\penalty0 1053--1062, 1999.

\bibitem[Dunkler et~al.(2014)Dunkler, Plischke, Leffondr{\'e}, and
  Heinze]{dunkler2014augmented}
D.~Dunkler, M.~Plischke, K.~Leffondr{\'e}, and G.~Heinze.
\newblock Augmented backward elimination: A pragmatic and purposeful way to
  develop statistical models.
\newblock \emph{PLOS ONE}, 9\penalty0 (11):\penalty0 1--19, 11 2014.
\newblock \doi{10.1371/journal.pone.0113677}.

\bibitem[Fogarty et~al.(2017)Fogarty, Shi, Mikkelsen, and
  Small]{fogarty2017randomization}
C.~B. Fogarty, P.~Shi, M.~E. Mikkelsen, and D.~S. Small.
\newblock Randomization inference and sensitivity analysis for composite null
  hypotheses with binary outcomes in matched observational studies.
\newblock \emph{Journal of the American Statistical Association}, 112\penalty0
  (517):\penalty0 321--331, 2017.

\bibitem[Greenland(2008)]{greenland2008invited}
S.~Greenland.
\newblock Invited commentary: Variable selection versus shrinkage in the
  control of multiple confounders.
\newblock \emph{American Journal of Epidemiology}, 167\penalty0 (5):\penalty0
  523--529, 2008.

\bibitem[Greenland et~al.(1999)Greenland, Robins, and
  Pearl]{greenland1999confounding}
S.~Greenland, J.~M. Robins, and J.~Pearl.
\newblock Confounding and collapsibility in causal inference.
\newblock \emph{Statistical Science}, 14\penalty0 (1):\penalty0 29--46, 1999.

\bibitem[Greenland et~al.(2016)Greenland, Daniel, and
  Pearce]{greenland2016outcome}
S.~Greenland, R.~Daniel, and N.~Pearce.
\newblock Outcome modelling strategies in epidemiology: traditional methods and
  basic alternatives.
\newblock \emph{International Journal of Epidemiology}, 45\penalty0
  (2):\penalty0 565--575, 2016.

\bibitem[H{\"a}ggstr{\"o}m(2019)]{haggstrom2019covsel}
J.~H{\"a}ggstr{\"o}m.
\newblock Personal communication, August 2019.

\bibitem[H{\"a}ggstr{\"o}m et~al.(2015)H{\"a}ggstr{\"o}m, Persson, Waernbaum,
  and de~Luna]{haggstrom2015covsel}
J.~H{\"a}ggstr{\"o}m, E.~Persson, I.~Waernbaum, and X.~de~Luna.
\newblock Covsel: An {R} package for covariate selection when estimating
  average causal effects.
\newblock \emph{Journal of Statistical Software}, 68\penalty0 (1):\penalty0
  1--20, 2015.

\bibitem[Hastie and Tibshirani(1990)]{hastie1990generalized}
T.~J. Hastie and R.~J. Tibshirani.
\newblock \emph{Generalized additive models}, volume~43.
\newblock CRC press, 1990.

\bibitem[Hoaglin(2016)]{hoaglin2016misunderstandings}
D.~C. Hoaglin.
\newblock Misunderstandings about q and `cochran's q test'in meta-analysis.
\newblock \emph{Statistics in Medicine}, 35\penalty0 (4):\penalty0 485--495,
  2016.

\bibitem[Imbens and Rubin(2015)]{imbens2015causal}
G.~W. Imbens and D.~B. Rubin.
\newblock \emph{Causal inference in statistics, social, and biomedical
  sciences}.
\newblock Cambridge University Press, 2015.

\bibitem[Ju et~al.(2019)Ju, Wyss, Franklin, Schneeweiss, H{\"a}ggstr{\"o}m, and
  van~der Laan]{cheng2019collaborative}
C.~Ju, R.~Wyss, J.~M. Franklin, S.~Schneeweiss, J.~H{\"a}ggstr{\"o}m, and M.~J.
  van~der Laan.
\newblock Collaborative-controlled lasso for constructing propensity
  score-based estimators in high-dimensional data.
\newblock \emph{Statistical Methods in Medical Research}, 28\penalty0
  (4):\penalty0 1044--1063, 2019.
\newblock \doi{10.1177/0962280217744588}.
\newblock URL \url{https://doi.org/10.1177/0962280217744588}.
\newblock PMID: 29226777.

\bibitem[Keele and Small(2018)]{keele2018comparing}
L.~Keele and D.~Small.
\newblock Comparing covariate prioritization via matching to machine learning
  methods for causal inference using five empirical applications.
\newblock \emph{arXiv preprint arXiv:1805.03743}, 2018.

\bibitem[Kim and Zambom(2019)]{kim2019signifreg}
J.~Kim and A.~Z. Zambom.
\newblock \emph{{SignifReg}: Consistent Significance Controlled Variable
  Selection in Linear Regression}, 2019.
\newblock URL \url{https://CRAN.R-project.org/package=SignifReg}.
\newblock R package version 2.1.

\bibitem[Kursa and Rudnicki(2010)]{kursa2010feature}
M.~B. Kursa and W.~R. Rudnicki.
\newblock Feature selection with the boruta package.
\newblock \emph{Journal of Statistical Software}, 36\penalty0 (11):\penalty0
  1--13, 2010.

\bibitem[LaLonde(1986)]{lalonde1986evaluating}
R.~J. LaLonde.
\newblock Evaluating the econometric evaluations of training programs with
  experimental data.
\newblock \emph{The American Economic Review}, 76\penalty0 (4):\penalty0
  604--620, 1986.

\bibitem[Maldonado and Greenland(1993)]{maldonado1993simulation}
G.~Maldonado and S.~Greenland.
\newblock Simulation study of confounder-selection strategies.
\newblock \emph{American journal of epidemiology}, 138\penalty0 (11):\penalty0
  923--936, 1993.

\bibitem[Mickey and Greenland(1989)]{mickey1989impact}
R.~M. Mickey and S.~Greenland.
\newblock The impact of confounder selection criteria on effect estimation.
\newblock \emph{American journal of epidemiology}, 129\penalty0 (1):\penalty0
  125--137, 1989.

\bibitem[Morgan and Rubin(2012)]{morgan2012rerandomization}
K.~L. Morgan and D.~B. Rubin.
\newblock Rerandomization to improve covariate balance in experiments.
\newblock \emph{The Annals of Statistics}, 40\penalty0 (2):\penalty0
  1263--1282, 2012.
\newblock \doi{10.1214/12-AOS1008}.

\bibitem[Neyman et~al.(1990)Neyman, Dabrowska, and Speed]{neyman1923}
J.~Neyman, D.~Dabrowska, and T.~Speed.
\newblock On the application of probability theory to agricultural experiments.
  {E}ssay on principles. {S}ection 9.
\newblock \emph{Statistical Science}, 5\penalty0 (4):\penalty0 465--472, 1990.

\bibitem[Pearl(2009)]{pearl2009}
J.~Pearl.
\newblock \emph{Causality: Models, Reasoning and Inference}.
\newblock Cambridge University Press, New York, NY, USA, 2nd edition, 2009.
\newblock \doi{10.1017/CBO9780511803161}.

\bibitem[Rosenbaum(1987)]{rosenbaum1987model}
P.~R. Rosenbaum.
\newblock Model-based direct adjustment.
\newblock \emph{Journal of the American Statistical Association}, 82\penalty0
  (398):\penalty0 387--394, 1987.

\bibitem[Rosenbaum(2002)]{Rosenbaum:2002}
P.~R. Rosenbaum.
\newblock \emph{Observational {S}tudies}.
\newblock New York : Springer, New York, 2002.

\bibitem[Rosenbaum and Rubin(1983)]{rosenbaum1983central}
P.~R. Rosenbaum and D.~B. Rubin.
\newblock The central role of the propensity score in observational studies for
  causal effects.
\newblock \emph{Biometrika}, 70\penalty0 (1):\penalty0 41--55, 1983.

\bibitem[Rubin(2007)]{rubin2007design}
D.~B. Rubin.
\newblock The design versus the analysis of observational studies for causal
  effects: parallels with the design of randomized trials.
\newblock \emph{Statistics in Medicine}, 26\penalty0 (1):\penalty0 20--36,
  2007.

\bibitem[Shortreed and Ertefaie(2017)]{shortreed2017outcome}
S.~M. Shortreed and A.~Ertefaie.
\newblock Outcome-adaptive lasso: Variable selection for causal inference.
\newblock \emph{Biometrics}, 73\penalty0 (4):\penalty0 1111--1122, 2017.

\bibitem[Speed(1992)]{Speed1992}
T.~P. Speed.
\newblock \emph{Introduction to Fisher (1926) The Arrangement of Field
  Experiments}, pages 71--81.
\newblock Springer Series in Statistics (Perspectives in Statistics). Springer,
  New York, NY, 1992.
\newblock ISBN 978-1-4612-4380-9.
\newblock \doi{10.1007/978-1-4612-4380-9_7}.
\newblock URL \url{https://doi.org/10.1007/978-1-4612-4380-9_7}.

\bibitem[Stephens et~al.(2013)Stephens, Tchetgen~Tchetgen, and
  De~Gruttola]{stephens2013flexible}
A.~J. Stephens, E.~J. Tchetgen~Tchetgen, and V.~De~Gruttola.
\newblock Flexible covariate-adjusted exact tests of randomized treatment
  effects with application to a trial of {HIV} education.
\newblock \emph{The Annals of Applied Statistics}, 7\penalty0 (4):\penalty0
  2106, 2013.

\bibitem[Stuart(2010)]{stuart2010matching}
E.~A. Stuart.
\newblock Matching methods for causal inference: A review and a look forward.
\newblock \emph{Statistical science: a review journal of the Institute of
  Mathematical Statistics}, 25\penalty0 (1):\penalty0 1, 2010.

\bibitem[Talbot et~al.(2015)Talbot, Lefebvre, and Atherton]{talbot2015bayesian}
D.~Talbot, G.~Lefebvre, and J.~Atherton.
\newblock The {B}ayesian causal effect estimation algorithm.
\newblock \emph{Journal of Causal Inference}, 3\penalty0 (2):\penalty0
  207--236, 2015.

\bibitem[Talbot et~al.(2017)Talbot, Lefebvre, Atherton, and
  Chiu.]{talbot2017bcee}
D.~Talbot, G.~Lefebvre, J.~Atherton, and Y.~Chiu.
\newblock \emph{BCEE: The Bayesian Causal Effect Estimation Algorithm}, 2017.
\newblock URL \url{https://CRAN.R-project.org/package=BCEE}.
\newblock R package version 1.2.

\bibitem[{van der Laan} and Gruber(2010)]{van2010collaborative}
M.~J. {van der Laan} and S.~Gruber.
\newblock Collaborative double robust targeted maximum likelihood estimation.
\newblock \emph{The international journal of biostatistics}, 6\penalty0 (1),
  2010.

\bibitem[VanderWeele(2019)]{vanderweele2019principles}
T.~J. VanderWeele.
\newblock Principles of confounder selection.
\newblock \emph{European journal of epidemiology}, 34\penalty0 (3):\penalty0
  211--219, 2019.

\bibitem[Vansteelandt and Keiding(2011)]{vansteelandt2011invited}
S.~Vansteelandt and N.~Keiding.
\newblock {Invited Commentary: G-Computation--Lost in Translation?}
\newblock \emph{American Journal of Epidemiology}, 173\penalty0 (7):\penalty0
  739--742, 03 2011.
\newblock ISSN 0002-9262.
\newblock \doi{10.1093/aje/kwq474}.

\bibitem[Vansteelandt et~al.(2012)Vansteelandt, Bekaert, and
  Claeskens]{vansteelandt2012model}
S.~Vansteelandt, M.~Bekaert, and G.~Claeskens.
\newblock On model selection and model misspecification in causal inference.
\newblock \emph{Statistical Methods in Medical Research}, 21\penalty0
  (1):\penalty0 7--30, 2012.

\bibitem[Wang et~al.(2015)Wang, Dominici, Parmigiani, and
  Zigler]{wang2015accounting}
C.~Wang, F.~Dominici, G.~Parmigiani, and C.~M. Zigler.
\newblock Accounting for uncertainty in confounder and effect modifier
  selection when estimating average causal effects in generalized linear
  models.
\newblock \emph{Biometrics}, 71\penalty0 (3):\penalty0 654--665, 2015.

\bibitem[Wilson and Reich(2014)]{wilson2014confounder}
A.~Wilson and B.~J. Reich.
\newblock Confounder selection via penalized credible regions.
\newblock \emph{Biometrics}, 70\penalty0 (4):\penalty0 852--861, 2014.

\bibitem[Witte and Didelez(2018)]{witte2018covariate}
J.~Witte and V.~Didelez.
\newblock Covariate selection strategies for causal inference: Classification
  and comparison.
\newblock \emph{Biometrical Journal}, 2018.

\bibitem[Wright and Ziegler(2017)]{rangerR2017}
M.~N. Wright and A.~Ziegler.
\newblock {ranger}: A fast implementation of random forests for high
  dimensional data in {C++} and {R}.
\newblock \emph{Journal of Statistical Software}, 77\penalty0 (1):\penalty0
  1--17, 2017.
\newblock \doi{10.18637/jss.v077.i01}.

\bibitem[Zambom and Kim(2018)]{zambom2018consistent}
A.~Z. Zambom and J.~Kim.
\newblock Consistent significance controlled variable selection in
  high-dimensional regression.
\newblock \emph{Stat}, 7\penalty0 (1):\penalty0 e210, 2018.
\newblock \doi{10.1002/sta4.210}.

\bibitem[Zubizarreta(2012)]{zubizarreta2012using}
J.~R. Zubizarreta.
\newblock Using mixed integer programming for matching in an observational
  study of kidney failure after surgery.
\newblock \emph{Journal of the American Statistical Association}, 107\penalty0
  (500):\penalty0 1360--1371, 2012.

\end{thebibliography}

\appendix

\section{Effect estimator assuming linear treatment and outcome models}\label{sect:est_linear}

In this section, we consider settings where linear treatment and outcome models can be reasonably assumed, by setting $g(\cdot)$ and $h(\cdot)$ as the identity link in \eqref{eq:ds_PSmodel} and \eqref{eq:ds_Ymodel} respectively. 
Under linear models for both treatment and outcome in the $j$-th orbit, respectively $\E(A|L^{j})=\alpha_{j} L^{j}$ and $\E(Y|A,L^{j})=\psi_{j} A + \beta_{j} L^{j}$, the ordinary least squares (OLS) estimator of the treatment effect $\widehat \psi_{j}$ in the $j$-th orbit obeys the following asymptotic expansion around $\psi$:
\beq\label{eq:ols_asymptotic}
n^{1/2}\left(\widehat\psi_{j}-\psi\right) = n^{-1/2} \sum_i
\dfrac{\left(A_i-\alpha_{j} L^{j}_i\right)\left(Y_i-\psi_{j} A_i - \beta_{j} L^{j}_i\right)}
{\E\left\{A(A-\alpha_{j} L^{j})\right\}}
+ o_p(1).
\eeq
The estimator is unbiased so long as either the treatment or outcome model is correctly specified.
It follows from the above expansion that the influence function for individual $i$ is:
\[
\phi^{j}_i = \dfrac{\left(A_i-\alpha_{j} L^{j}_i\right)\left(Y_i-\psi_{j} A_i - \beta_{j} L^{j}_i\right)}
{\E\left\{A(A-\alpha_{j} L^{j})\right\}}.
\]
Let $\widehat \phi^{j}_i = {\left(A_i-\widehat\alpha_{j} L^{j}_i\right)\left(Y_i-\widehat\psi_{j} A_i - \widehat\beta_{j} L^{j}_i\right)}\left/
{n^{-1} \sum_i A_i\left(A_i-\widehat\alpha_{j} L^{j}_i\right)}\right.$ denote the estimated influence function, obtained by plugging in the OLS estimators $\widehat\alpha_{j},\widehat\psi_{j},\widehat\beta_{j}$ for the coefficients in the treatment and outcome models, and substituting the population expectation with a sample average.
The variance of the difference between effect estimators from two different orbits, e.g., $j$ and $k$, is consistently estimated by the sample variance of the corresponding difference in estimated influence functions:
\beq\label{eq:var_diff_infn}
\widehat\V\left\{n^{1/2}\left(\widehat\psi_{j}-\widehat\psi_{k}\right)\right\}
=
(n-1)^{-1} \sum_i \left(\widehat\phi^{j}_i - \widehat\phi^{k}_i\right)^2.
\eeq
It follows that under settings with linear regression models for the treatment and the outcome, consistency and asymptotic normality of the standardized difference \eqref{eq:score_diff_std_fullmodel} with mean zero and variance one directly follow from the law of large numbers and the central limit theorem.

\section{Simulation study results}\label{sect:sim_results:appx}

\begin{landscape}

\begin{table}[ht]
\centering
\setlength{\tabcolsep}{3pt}
\begin{tabular}{cccc|rrrrr|rrrrr|rrrrrrr}
  \hline
 & & & & \multicolumn{5}{c}{Prob. confounders selected} 
 & \multicolumn{5}{c}{No. selected covariates} & \multicolumn{7}{c}{Type I error rate ($\alpha=0.05$)} \\  
${\cal S}_3$ & $p$ & $|{\cal S}_3|$ & $Y$ 
& (A) & (B) & (C) & (SR) & (St) 
& (A) & (B) & (C) & (SR) & (St) & (No) & (A) & (B) & (C) & (SR) & (St) & (Ta) \\ 
  \hline
IV & 25 & 2 & Con. & 0.46 & 0.07 & 0.69 & 0.06 & 0.75 & 9.37 & 1.79 & 6.51 & 1.83 & 9.93 & 0.13 & 0.06 & 0.10 & 0.07 & 0.07 & 0.06 & 0.06 \\ 
IV & 25 & 2 & Bin. & 0.95 & 0.32 & 0.60 & 0.43 & 0.81 & 19.30 & 3.15 & 3.75 & 3.22 & 10.94 & 0.27 & 0.11 & 0.14 & 0.07 & 0.09 & 0.08 & 0.05 \\ 
IV & 25 & 4 & Con. & 0.49 & 0.07 & 0.58 & 0.10 & 0.50 & 9.49 & 1.84 & 7.74 & 1.97 & 9.28 & 0.10 & 0.05 & 0.08 & 0.04 & 0.05 & 0.06 & 0.04 \\ 
IV & 25 & 4 & Bin. & 0.95 & 0.31 & 0.48 & 0.44 & 0.69 & 19.46 & 3.07 & 4.12 & 3.20 & 11.17 & 0.19 & 0.11 & 0.10 & 0.08 & 0.06 & 0.09 & 0.05 \\ 
IV & 60 & 2 & Con. & 0.52 & 0.04 & 0.02 & 0.08 & 0.67 & 36.95 & 2.36 & 0.40 & 2.78 & 16.26 & 0.12 & 0.10 & 0.09 & 0.12 & 0.08 & 0.08 & 0.05 \\ 
IV & 60 & 2 & Bin. & 0.98 & 0.23 & 0.28 & 0.37 & 0.91 & 58.20 & 3.48 & 10.45 & 4.03 & 33.98 & 0.24 & 0.23 & 0.13 & 0.21 & 0.07 & 0.21 & 0.05 \\ 
IV & 60 & 4 & Con. & 0.54 & 0.04 & 0.02 & 0.07 & 0.46 & 36.51 & 2.35 & 0.44 & 2.80 & 16.33 & 0.10 & 0.09 & 0.08 & 0.10 & 0.05 & 0.06 & 0.05 \\ 
IV & 60 & 4 & Bin. & 0.97 & 0.23 & 0.23 & 0.40 & 0.84 & 58.42 & 3.48 & 10.05 & 4.03 & 33.85 & 0.20 & 0.17 & 0.12 & 0.16 & 0.06 & 0.16 & 0.04 \\ 
Co. & 25 & 2 & Con. & 0.49 & 0.06 & 0.82 & 0.07 & 0.95 & 9.21 & 1.75 & 5.32 & 1.86 & 10.70 & 0.19 & 0.08 & 0.14 & 0.07 & 0.10 & 0.06 & 0.06 \\ 
Co. & 25 & 2 & Bin. & 0.95 & 0.28 & 0.72 & 0.37 & 0.96 & 18.43 & 3.07 & 3.41 & 3.11 & 10.50 & 0.45 & 0.07 & 0.20 & 0.09 & 0.12 & 0.05 & 0.05 \\ 
Co. & 25 & 4 & Con. & 0.46 & 0.06 & 0.80 & 0.07 & 0.95 & 9.18 & 1.80 & 5.30 & 1.81 & 10.55 & 0.19 & 0.06 & 0.15 & 0.07 & 0.10 & 0.06 & 0.06 \\ 
Co. & 25 & 4 & Bin. & 0.94 & 0.30 & 0.70 & 0.38 & 0.97 & 18.58 & 3.13 & 3.43 & 3.13 & 10.71 & 0.47 & 0.07 & 0.19 & 0.09 & 0.12 & 0.05 & 0.06 \\ 
Co. & 60 & 2 & Con. & 0.52 & 0.04 & 0.04 & 0.06 & 0.85 & 36.54 & 2.40 & 0.68 & 2.72 & 13.38 & 0.18 & 0.12 & 0.12 & 0.17 & 0.09 & 0.08 & 0.05 \\ 
Co. & 60 & 2 & Bin. & 0.96 & 0.22 & 0.36 & 0.32 & 0.95 & 58.04 & 3.50 & 10.66 & 3.86 & 31.37 & 0.41 & 0.37 & 0.20 & 0.28 & 0.11 & 0.28 & 0.04 \\ 
Co. & 60 & 4 & Con. & 0.53 & 0.03 & 0.05 & 0.06 & 0.86 & 36.58 & 2.28 & 0.77 & 2.75 & 13.31 & 0.17 & 0.12 & 0.14 & 0.16 & 0.10 & 0.07 & 0.05 \\ 
Co. & 60 & 4 & Bin. & 0.97 & 0.22 & 0.34 & 0.33 & 0.95 & 58.07 & 3.45 & 10.36 & 3.85 & 31.52 & 0.44 & 0.40 & 0.20 & 0.29 & 0.12 & 0.28 & 0.05 \\ 
  \hline
\end{tabular}
\caption{Results for each covariate selection method used to determine the PS model for full matching.
The methods were no covariates (`No'), ABE (A) , Boruta (B), CovSel (C), SignifReg (SR), Stability (St), or the known target covariates (Ta).
The empirical proportions (`Prob.') of simulated datasets where exactly both confounders were included among the selected covariates, 
the average number (`No.') of selected covariates, and the type I error rate of the randomization tests (significance level $\alpha=0.05$) were calculated.
The observed covariates indexed by $s \in {\cal S}_3$ either affected treatment only (`IV'), or induced collider bias when adjusted for (`Co').
There were $p$ observed covariates. The outcome $Y$ was either continuous (`Con.') or binary (`Bin.').
All results were rounded to two decimal places.
\label{table:sims-covsel-compare}}
\end{table}  

\begin{table}[ht]
\centering
\setlength{\tabcolsep}{5pt}
\begin{tabular}{cccc|rrr|rrr|rrr|rrr|rrr}
  \hline
 & & & & \multicolumn{3}{c}{Type I} 
 & \multicolumn{3}{c}{Mean effect est.} & \multicolumn{3}{c}{E.s.e} 
 & \multicolumn{3}{c}{Mean s.e. est.}  & \multicolumn{3}{c}{A.s.e}  \\
${\cal S}_3$ & $p$ & $|{\cal S}_3|$ & $Y$ 
& (i) & (ii) & (iii) & (i) & (ii) & (iii) & (i) & (ii) & (iii) & (i) & (ii) & (iii) & (i) & (ii) & (iii) \\ 
  \hline
IV & 25 & 2 & Con. & 0.60 & 0.43 & 0.11 & 1.07 & 0.98 & 0.64 & 2.71 & 3.77 & 1.28 & 0.68 & 0.82 & 1.21 & 0.20 & 0.30 & 0.19 \\ 
IV & 25 & 2 & Bin. & 0.66 & 0.60 & 0.12 & 0.17 & 0.11 & 0.07 & 0.26 & 0.30 & 0.14 & 0.07 & 0.10 & 0.13 & 0.02 & 0.36 & 0.02 \\ 
IV & 25 & 4 & Con. & 0.59 & 0.48 & 0.08 & 1.02 & 0.84 & 0.60 & 2.58 & 4.40 & 1.23 & 0.60 & 0.86 & 1.23 & 0.14 & 2.72 & 0.21 \\ 
IV & 25 & 4 & Bin. & 0.65 & 0.62 & 0.09 & 0.14 & 0.08 & 0.07 & 0.24 & 0.32 & 0.14 & 0.07 & 0.10 & 0.13 & 0.05 & 0.39 & 0.02 \\ 
Co. & 25 & 2 & Con. & 0.41 & 0.35 & 0.08 & 0.36 & 0.61 & 0.32 & 2.38 & 2.24 & 1.13 & 0.92 & 0.91 & 1.07 & 0.30 & 0.67 & 0.14 \\ 
Co. & 25 & 2 & Bin. & 0.35 & 0.46 & 0.08 & 0.07 & 0.06 & 0.03 & 0.24 & 0.22 & 0.13 & 0.10 & 0.11 & 0.12 & 0.03 & 0.44 & 0.01 \\ 
Co. & 25 & 4 & Con. & 0.40 & 0.36 & 0.08 & 0.30 & 0.38 & 0.33 & 2.43 & 2.02 & 1.10 & 0.92 & 0.91 & 1.06 & 0.27 & 0.50 & 0.14 \\ 
Co. & 25 & 4 & Bin. & 0.34 & 0.46 & 0.09 & 0.07 & 0.07 & 0.05 & 0.24 & 0.22 & 0.12 & 0.10 & 0.10 & 0.12 & 0.03 & 0.05 & 0.01 \\ 
  \hline    
\end{tabular}
\caption{Empirical type I error rate ($\alpha=0.05$) of the p-values for testing $H_0$, estimates (est.) of the treatment effect, and its standard error (s.e.), using the CTMLE and HDM methods. 
The methods were (i) CTMLE: LASSO, (ii) CTMLE: discrete, or (iii) HDM.
The empirical s.e. (E.s.e.) of the effect estimates, and asymptotic s.e. (A.s.e.) of the standard error estimates (i.e., its standard deviation across the simulated datasets), were considered. 
The observed covariates indexed by $s \in {\cal S}_3$ either affected treatment only (`IV'), or induced collider bias when adjusted for (`Co').
There were $p$ observed covariates. The outcome $Y$ was either continuous (`Con.') or binary (`Bin.').
The true value of the treatment effect was zero.
All results were rounded to two decimal places.
\label{table:sims-ATE-compare-appx}}
\end{table}  

\end{landscape}

Two Bayesian methods for treatment effect estimation with covariate selection were also employed. 
For the Bayesian Adjustment for Confounding (BAC) algorithm, as implemented in the \texttt{bac} function in the \texttt{bacr} package \citep{wang2015accounting}, the number of MCMC iterations was set to 5000, with a burn-in of 500, and a thinning parameter of 10. For the Bayesian causal effect estimation algorithm \citep{talbot2015bayesian}, as implemented in the \texttt{ABCEE} function in the \texttt{BCEE} package \citep{talbot2017bcee}, default levels of all options were used. The argument \texttt{omega} was set to the recommended value of $500\sqrt{n}$. 
For each method, the posterior probability that the estimated treatment effect was greater than zero was used as a p-value for testing $H_0$. 
The results for two selected settings are plotted in Figure~\ref{fig:sims-others2-covsel} below.

    \begin{figure}[!ht]
    \centering
\includegraphics[width=.49\linewidth,keepaspectratio]{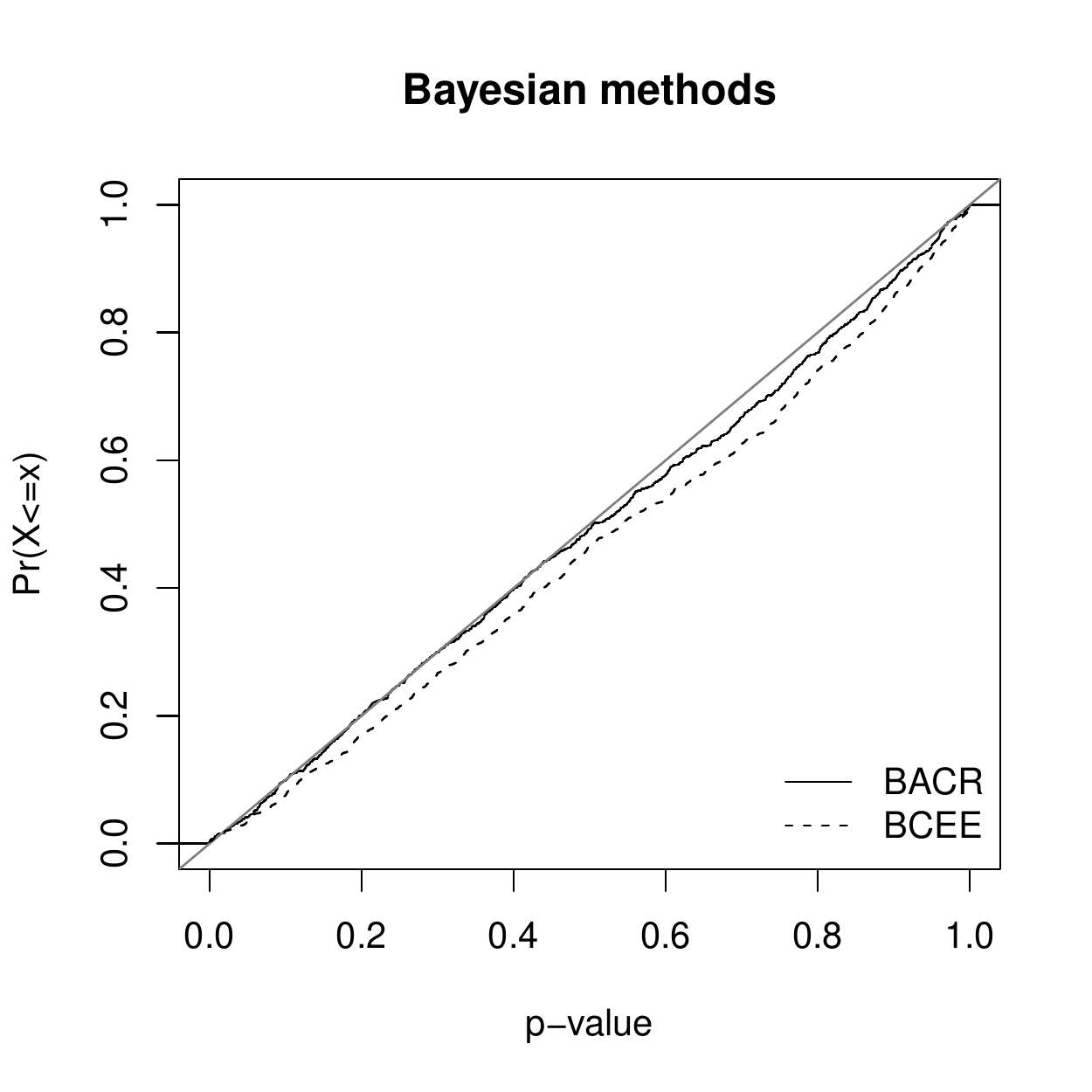}
\includegraphics[width=.49\linewidth,keepaspectratio]{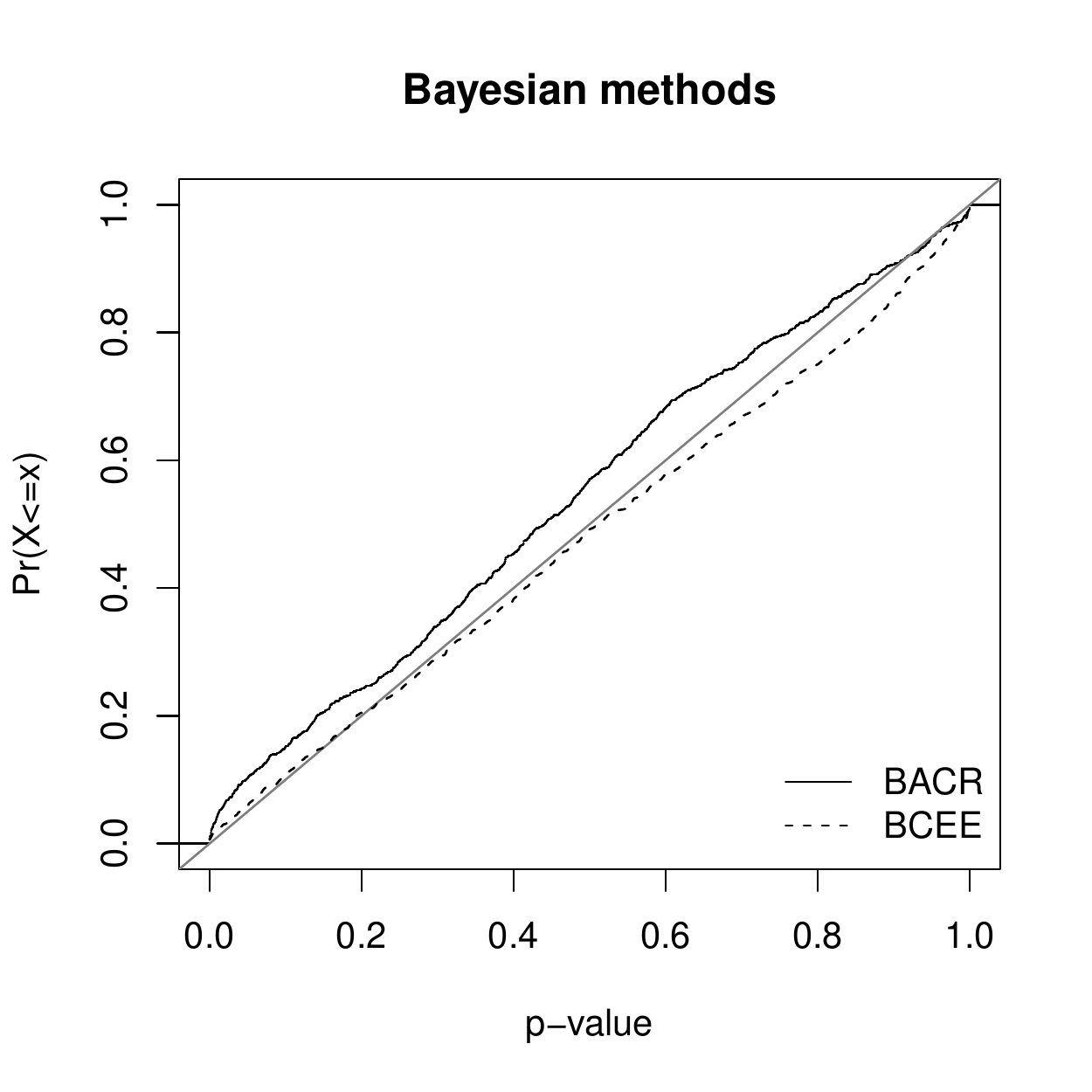}
    \caption{Empirical distributions of p-values for testing $H_0$ using Bayesian approaches for causal effect estimation.
    Observed covariates indexed by $s \in {\cal S}_3 = \{5, 6\}$ either affected treatment only (left panel), or induced collider bias when adjusted for (right panel).
    There were $p=25$ covariates, and the outcome was continuous.
    The random variable $X$ on the vertical axes denote the p-value.     
    The diagonal is plotted as a solid (gray) line with reduced opacity.
    \label{fig:sims-others2-covsel}}
    \end{figure}

\clearpage

\section{Selected covariates in the RHC data}

\begin{longtable}{ |p{3in}|c|c|c|c|c|c| } 
 \hline
 Covariates & Stable & ABE & Boruta & SignifReg & All \\
 \hline\endhead
Cancer: Metastatic (ca\_meta) & TRUE & TRUE & TRUE & TRUE & TRUE \\ 
  Support model estimate of the prob. of surviving 2 months (surv2md1) & TRUE & TRUE & TRUE &  & TRUE \\ 
  Heart rate (hrt1) & TRUE & TRUE &  & TRUE & TRUE \\ 
  Secondary disease: Colon Cancer (cat2\_colon) & TRUE & TRUE &  &  & TRUE \\ 
  Orthopedic Diagnosis (ortho) & TRUE & TRUE & TRUE & TRUE & TRUE \\ 
  Temperature (temp1) & TRUE &  &  &  & TRUE \\ 
  Mean blood pressure (meanbp1) & TRUE &  &  &  & TRUE \\ 
  Duke Activity Status Index (das2d3pc) & TRUE & TRUE &  & TRUE & TRUE \\ 
  Cardiovascular Diagnosis (card) & TRUE & TRUE &  &  & TRUE \\ 
  Primary disease: MOSF with Sepsis (cat1\_mosfsep) & TRUE &  & TRUE &  & TRUE \\ 
  Primary disease: CHF (cat1\_chf) & TRUE &  & TRUE &  & TRUE \\ 
  PaO2/FIO2 ratio (pafi1) & TRUE &  &  &  & TRUE \\ 
  Albumin (alb1) & TRUE & TRUE & TRUE & TRUE & TRUE \\ 
  Secondary disease category: MOSF with Sepsis (cat2\_mosfsep) & TRUE &  &  &  & TRUE \\ 
  Immunosupperssion, Organ Transplant, HIV Positivity, Diabetes Mellitus Without End Organ Damage, Diabetes Mellitus With End Organ Damage, Connective Tissue Disease (immunhx) & TRUE &  &  &  & TRUE \\ 
  Hematocrit (hema1) & TRUE & TRUE &  &  & TRUE \\ 
  Primary disease category: Cirrhosis (cat1\_cirr) & TRUE &  &  &  & TRUE \\ 
  Weight (wtkilo1) & TRUE & TRUE & TRUE & TRUE & TRUE \\ 
  Upper GI Bleeding (gibledhx) & TRUE &  &  &  & TRUE \\ 
  Urine output (urin1) & TRUE & TRUE & TRUE & TRUE & TRUE \\ 
  Sepsis Diagnosis (seps) & TRUE &  &  &  & TRUE \\ 
  Glasgow Coma Score (scoma1) & TRUE & TRUE & TRUE &  & TRUE \\ 
  Definite Myocardial Infarction (amihx) & TRUE & TRUE & TRUE &  & TRUE \\ 
  Sodium (sod1) & TRUE &  & TRUE &  & TRUE \\ 
  PaCo2 (paco21) & TRUE &  &  &  & TRUE \\ 
  Medical insurance: Medicaid (ins\_caid) & TRUE & TRUE &  &  & TRUE \\ 
  Respiratory Rate (resp1) & TRUE & TRUE & TRUE & TRUE & TRUE \\ 
  Respiratory Diagnosis (resp) & TRUE &  &  &  & TRUE \\ 
  White Blood Cell Count (wblc1) & TRUE &  &  &  & TRUE \\ 
  Creatinine (crea1) &  & TRUE &  & TRUE & TRUE \\ 
  Primary disease category: Coma (cat1\_coma) &  & TRUE &  &  & TRUE \\ 
  (age) &  & TRUE &  & TRUE & TRUE \\ 
  Primary disease category: MOSF with Malignancy (cat1\_mosfmal) &  & TRUE &  & TRUE & TRUE \\ 
  Secondary disease category: MOSF with Malignancy (cat2\_mosfmal) &  & TRUE &  & TRUE & TRUE \\ 
  Secondary disease category: Coma (cat2\_coma) &  &  & TRUE &  & TRUE \\ 
  Secondary disease category: Cirrhosis (cat2\_cirr) &  & TRUE &  &  & TRUE \\ 
  (sex) &  & TRUE &  &  & TRUE \\ 
  Solid Tumor, Metastatic Disease, Chronic Leukemia/Myeloma, Acute Leukemia, Lymphoma (malighx) &  & TRUE &  & TRUE & TRUE \\ 
  Income: 11-25k (income2) &  & TRUE &  &  & TRUE \\ 
  Acute MI, Peripheral Vascular Disease, Severe Cardiovascular Symptoms with NYHA-Class III, Very Severe Cardiovascular Symptoms with NYHA-Class IV (cardiohx) &  & TRUE &  &  & TRUE \\ 
  Dementia, Stroke or Cerebral Infact, Parkinson's Disease (dementhx) &  &  &  &  & TRUE \\ 
  Arterial PH (ph1) &  & TRUE &  &  & TRUE \\ 
  Metabolic Diagnosis (meta) &  & TRUE &  &  & TRUE \\ 
  Cancer: Yes (ca\_yes) &  &  & TRUE &  & TRUE \\ 
  Medical insurance: Medicare (ins\_care) &  & TRUE &  &  & TRUE \\ 
  Gastrointestinal Diagnosis (gastr) &  & TRUE &  &  & TRUE \\ 
  Primary disease category: COPD (cat1\_copd) &  &  &  &  & TRUE \\ 
  Years of education (edu) &  &  &  &  & TRUE \\ 
  Congestive Heart Failure (chfhx) &  &  &  &  & TRUE \\ 
  Chronic Renal Disease, Chronic Hemodialysis or Peritoneal Dialysis (renalhx) &  & TRUE &  &  & TRUE \\ 
  Income: Under 11k (income1) &  & TRUE &  &  & TRUE \\ 
  APACHE score (aps1) &  &  &  &  & TRUE \\ 
  Neurological Diagnosis (neuro) &  & TRUE &  &  & TRUE \\ 
  Hematologic Diagnosis (hema) &  &  &  &  & TRUE \\ 
  Potassium (pot1) &  &  & TRUE &  & TRUE \\ 
  Chronic Pulmonary Disease, Severe Pulmonary Disease, Very Severe Pulmonary Disease (chrpulhx) &  &  &  &  & TRUE \\ 
  Bilirubin (bili1) &  &  & TRUE &  & TRUE \\ 
  Trauma Diagnosis (trauma) &  &  & TRUE &  & TRUE \\ 
  Psychiatric History, Active Psychosis or Severe Depression (psychhx) &  &  &  &  & TRUE \\ 
  Medical insurance: None (ins\_no) &  &  &  &  & TRUE \\ 
  Race: Other (raceother) &  &  &  &  & TRUE \\ 
  Primary disease category: Colon Cancer (cat1\_colon) &  &  &  &  & TRUE \\ 
  Medical insurance: Medicare and Medicaid (ins\_carecaid) &  &  &  &  & TRUE \\ 
  Do Not Resuscitate (dnr1) &  &  & TRUE &  & TRUE \\ 
  Transfer from Another Hospital (transhx) &  &  &  &  & TRUE \\ 
  Secondary disease category: Lung Cancer (cat2\_lung) &  & TRUE &  &  & TRUE \\ 
  Cirrhosis, Hepatic Failure (liverhx) &  &  & TRUE &  & TRUE \\ 
  Medical insurance: Private (ins\_pcare) &  &  &  &  & TRUE \\ 
  Renal Diagnosis (renal) &  &  &  &  & TRUE \\ 
  Income: 25-50k (income3) &  &  & TRUE &  & TRUE \\ 
  Race: African American (raceblack) &  &  &  &  & TRUE \\ 
  Primary disease category: Lung Cancer (cat1\_lung) &  & TRUE &  &  & TRUE \\ 
 \hline
p-value for testing $H_0$ & 0.193 & 0.051 & 0.596 & 0.061 & 0.377 \\
Treatment effect estimate & 0.059 & 0.059 & 0.006 & 0.046 & 0.072 \\
Standard error & 0.022 & 0.019 & 0.019 & 0.018 & 0.022 \\
\hline

\caption{Comparison of the selected confounders (``TRUE'') in the RHC data using different methods. The covariates were ordered following double selection principles \eqref{eq:forwardselect_minimin}. 
The p-values for testing $H_0$, and estimated average (marginal) treatment effect and standard error, are stated (up to three decimal places) in the last three rows. There were no results for the CovSel covariate selection method as it could not be implemented for this dataset.
The results using a PS model with all the available covariates are shown in the rightmost column (`All').
\label{table:rhc-compare}}
\end{longtable}       

\end{document}